\documentclass[12pt]{article}

\usepackage{amsmath}
\usepackage{amssymb}

\newcommand{\al}{\alpha}
\newcommand{\ep}{\epsilon}

\newcommand{\la}{\lambda}

\newcommand{\lb}{\lbrack}
\newcommand{\rb}{\rbrack}

\newcommand{\msc}[1]{\mbox{\scriptsize #1}}
\newcommand{\dsp}{\displaystyle}

\newcommand{\br}{\Bbb R}
\newcommand{\bz}{\Bbb Z}

\newcommand{\bsz}{\Bbb Z}

\newcommand{\bl}{{\bf l}}
\newcommand{\tbl}{\tilde{\bl}}
\newcommand{\bk}{{\bf k}}

\newcommand{\cG}{{\cal G}}

\newcommand{\cT}{{\cal T}}

\newcommand{\cN}{{\cal N}}
\newcommand{\cM}{{\cal M}}

\newcommand{\cC}{{\cal C}}

\newcommand{\cD}{{\cal D}}

\newcommand{\cH}{{\cal H}}

\newcommand{\cK}{{\cal K}}

\newcommand{\tL}{\tilde{L}}

\newcommand{\tI}{\tilde{I}}

\newcommand{\tell}{\tilde{\ell}}

\newcommand{\tal}{\tilde{\al}}

\newcommand{\hZ}{\widehat{Z}}

\newcommand{\hsigma}{\widehat{\sigma}}

\newcommand{\ket}[1]{{|#1\rangle}}
\newcommand{\bra}[1]{{\langle#1|}}

\newcommand{\Th}[2]{\Theta_{#1,#2}}
\renewcommand{\th}{{\theta}}

\newcommand{\ch}[2]{\mbox{ch}^{#1}_{#2}}

\newcommand{\tr}{\mbox{Tr}}
\renewcommand{\mod}{\mbox{mod}}

\newcommand{\nn}{\nonumber\\}

\newcommand{\NS}{\mbox{NS}}
\newcommand{\tNS}{\widetilde{\mbox{NS}}}
\newcommand{\R}{\mbox{R}}
\newcommand{\tR}{\widetilde{\mbox{R}}}
\newcommand{\sNS}{\msc{NS}}
\newcommand{\stNS}{\widetilde{\msc{NS}}}
\newcommand{\sR}{\msc{R}}
\newcommand{\stR}{\widetilde{\msc{R}}}

\newcommand {\eqn}[1]{(\ref{#1})}

\makeatletter
\@addtoreset{equation}{section}
\def\theequation{\thesection.\arabic{equation}}
\makeatother

\begin{document}

\begin{titlepage}
 \
 \renewcommand{\thefootnote}{\fnsymbol{footnote}}
 \font\csc=cmcsc10 scaled\magstep1
 {\baselineskip=14pt
 \rightline{
 \vbox{\hbox{arXiv:0711.1045 [hep-th]}
       \hbox{UT-07-36}
       \hbox{HIP-07-59/TH}
       }}}

 \baselineskip=20pt
\vskip 2cm
 
\begin{center}

{\huge  Mirrorfolds with K3 Fibrations}



\vskip 1.0cm

{\large Shinsuke Kawai${}^{1}$}\footnote{{\tt shinsuke.kawai(AT)helsinki.fi}} and
{\large Yuji Sugawara${}^{2}$}\footnote{{\tt sugawara(AT)hep-th.phys.s.u-tokyo.ac.jp}}
\\[5mm]
${}^{1}${\it Helsinki Institute of Physics, 
P.O.Box 64, University of Helsinki, Helsinki 00014, Finland}\\
${}^{2}${\it Department of Physics, University of Tokyo, 7-3-1 Hongo, 
Bunkyo-ku, Tokyo 113-0033, Japan}
\\[15mm]

\end{center}

\vskip .5cm

\begin{abstract}

We study a class of non-geometric string vacua realized as completely
soluble superconformal field theory (SCFT). 
These models are defined as `interpolating orbifolds' of 
$K3 \times S^1$ by the mirror transformation acting on the $K3$ fiber 
combined with the half-shift on the $S^1$-base.
They are variants of the T-folds, the interpolating orbifolds 
by T-duality transformations, and thus may be called `mirrorfolds'. 
Starting with arbitrary (compact or non-compact) Gepner models 
for the $K3$ fiber, we construct modular invariant partition functions 
of general mirrorfold models.
In the case of compact $K3$ fiber the mirrorfolds only yield non-supersymmetric string vacua. 
They exhibit IR instability due to winding tachyon condensation which is
similar to the Scherk-Schwarz type circle compactification. 
When the fiber SCFT is non-compact (say, the ALE space in the simplest case), 
on the other hand, both supersymmetric and non-supersymmetric vacua can be constructed. 
The non-compact non-supersymmetric mirrorfolds 
can get stabilised at the level of 
string perturbation theory.
We also find that in the non-compact supersymmeric mirrorfolds D-branes 
are {\em always} non-BPS.
These D-branes can get stabilized  
against both open- and closed-string marginal deformations.  

\end{abstract}

\vfill

\setcounter{footnote}{0}
\renewcommand{\thefootnote}{\arabic{footnote}}
\end{titlepage}
\baselineskip 18pt


\section{Introduction and Summary}

String theory on non-geometric backgrounds has recently been receiving 
much attention. 
A particularly accessible class of non-geometric backgrounds
is those formulated as fibrations over a base manifold in which the transition 
functions are built from discrete duality transformations of string theory 
besides diffeomorphisms.
In such models the moduli space of the fibre, when going around non-trivial 
cycles on the base manifold, picks up monodromies in general; for this reason
these string vacua are often called `monodrofolds.' 
In particular, monodrofolds constructed from T-duality transformations are 
called `T-folds'\cite{Hull}.
Known examples of T-folds include those arising from flux-compactified type II strings
combined with T-duality. 
These are non-geometric in the sense that while they are locally equipped with 
geometric structures, globally they are not.
It is now increasingly recognised that such backgrounds constitute a natural and 
essential part of string vacua.
For recent topics and developments of non-geometric backgrounds in string theory, 
see e.g. \cite{Wecht-review} and references therein.

In studying non-geometric backgrounds that do not necessarily allow
intuitive geometric picture, approach by world-sheet conformal field theory (CFT) 
proves to be extremely powerful.
From CFT one may extract various essential information.
Firstly, consistency of the string vacua can be 
examined through modular invariance, 
locality of vertex operators, etc.
One may also find spectra of physical excitations, presence/absence of space-time 
supersymmetry (SUSY), as well as stability of the system.
While limited to the lowest order in the string coupling expansion, 
CFT gives all-order results in the $\al'$-correction beyond the supergravity approximation.
By now, several models of T-folds have been analysed using CFT\cite{FloW,HelW,doubled}.
Detailed study of D-branes in simple T-fold models
was also carried out by the present authors in \cite{KawaiS1}, where consistent D-branes on 
these backgrounds are explicitly constructed in boundary CFT, supporting and supplementing
previous observations of \cite{Lawrence:2006ma}.

In CFT, T-folds are typically realised as asymmetric interpolating orbifolds. 
They provide interesting models of string vacua as they generally involve less moduli.
Moreover, construction of such CFT models is delicate in general (e.g. achieving modular 
invariance), giving rise to stringent consistency checks. 
One may also hope for breaking SUSY while keeping 
attractive features of SUSY intact in 
such models, as discussed in \cite{Kachru:1998hd} based on toroidal models.
In the present article we apply 
techniques of interpolating orbifold CFT to more non-trivial backgrounds
of superstring theory.
The models we shall study are $K3$ fibrations over an $S^1$ base with the mirror twist, which
we call `mirrorfolds,' following the precedent examples of the monodrofolds and T-folds.
These are modelled in CFT as interpolating orbifolds of $K3 \times S^1$ with the mirror involution 
acting on the $K3$ fiber, which may be seen as extensions of the simplest T-folds mentioned above.
Note that such orbifolds are possible since $K3$ is self-dual for the mirror symmetry.
We shall see that the CFT machinery works well for these non-trivial {\em curved\/} fiber spaces. 
Similar models of string theory compactification involving $K3$ twists are investigated also in 
\cite{CLS}.

~


Main outcomes of this paper are summarized as follows: 

~

\noindent
{\bf 1. } 
We start by considering an arbitrary Gepner model \cite{Gepner}
to describe the $K3$ fiber. 
Besides the standard Gepner models for compact spaces we also treat 
non-compact models in which gravity decouples
\cite{ncGepner,ES-BH,ES-C}.\footnote
    {A recent study of the non-compact Gepner-like models has been
    given in \cite{Ashok:2007ui}.}
We elaborate on the construction of modular invariant partition functions in full generality. 
Careful fixing of phase ambiguity that appears in the mirror involution turns out to be crucial 
for the modular invariance. 

~

\noindent
{\bf 2.} 
In the case of the compact $K3$ fiber, the mirrorfolds yield only
non-SUSY string vacua.
They exhibit IR instability caused by winding tachyon condensation which is 
similar to the Scherk-Schwarz type circle compactification \cite{SS}. 

~

\noindent
{\bf 3.}
In the case of the non-compact $K3$ fiber (e.g. the ALE spaces), 
both SUSY and non-SUSY vacua can be constructed. 
The non-compact, non-SUSY mirrorfolds can be stabilised 
at the level of perturbative string. 
Namely, they get stable against arbitrary marginal deformations
of normalizable modes. 


~

\noindent
{\bf 4.}
The vacua of non-compact SUSY mirrorfolds are stable, of course. 
However, once putting an arbitrary consistent D-brane on these backgrounds, 
the space-time SUSY is inevitably broken. 
We examine the stability of such non-SUSY vacua, 
and find that 
the vacuum may become free from instability caused by open string tachyons.


~

This paper is organized as follows.
In section 2, starting with a brief review on the Gepner
construction of $K3$, we discuss 
the construction of modular invariant partition 
functions describing string theory
on the mirrorfolds with compact $K3$ fibrations. 
In section 3, we study the models with non-compact fibrations.
There are several common features in the compact and non-compact 
mirrorfolds, but there are also remarkable differences.
In section 4, we present discussions and outlook for
future work. 
In the Appendices we summarize
our notations of modular functions and various character formulas 
appearing in the main text.

We use the convention of $\al'=1$ throughout this paper. 

~


\section{Mirrorfolds with $K3$ Fibrations}

The superconformal system which we focus on in this paper is 
the {\em interpolating orbifolds} of the type
\begin{eqnarray}
 \frac{K3 \times S^1_{2R}}{\sigma_{\msc{mirror}} \otimes 
\cT_{2\pi R}},
\label{mirrorfold}
\end{eqnarray}
where $\sigma_{\msc{mirror}}$ 
denotes the mirror involution acting on the $K3$ `fiber', and 
$\cT_{2\pi R}$ 
denotes the half-shift (let the radius of $S^1$ be $2R$)
along the `base' $S^1$-direction:
\begin{eqnarray}
 && \cT_{2\pi R} ~:~ Y\, \longmapsto \, Y+2\pi R ~.
\label{half shift} 
\end{eqnarray}
This conformal system is expected to describe the $K3$-fibration 
over the $S^1$-base of radius $R$ (reduced by the half-shift 
$\cT_{2\pi R}$),
twisted by the mirror-transformation on $K3$. 
We assume an arbitrary Gepner model for the $K3$ fiber. 
We will also work with non-compact models in which gravity decouples, 
in the next section. 

~

\subsection{Preliminary : Gepner Models for $K3$}

In order to establish notations we start with 
a brief review of the Gepner construction of 
$K3$:
\begin{eqnarray}
 && \left\lb M_{k_1}\otimes \cdots  \otimes
M_{k_r}\right\rb\left|_{\bsz_N\msc{-orbifold}}
\right. ~, ~~~ \sum_{i=1}^r \frac{k_i}{k_i+2}=2~,
\label{Gepner K3}
\end{eqnarray}
where $M_k$ denotes the $\cN=2$ minimal model of level $k$
($\hat{c}\equiv \frac{c}{3}= \frac{k}{k+2}$), and we 
set
\begin{eqnarray}
 N \equiv \mbox{L.C.M.} \{k_i+2~;~i=1,\ldots,r\}~.
\end{eqnarray}
The $\bz_N$-orbifold means to project the Hilbert space 
onto the subspace with integer $U(1)_R$-charge. 
To keep consistency of the conformal field theory,  
this projection has to be accompanied by 
the twisted sectors generated by integral spectral flows \cite{EOTY}. 


The modular invariant of the model \eqn{Gepner K3} generically
has the following form (assuming the diagonal modular invariant 
with respect to the spin structure):
\begin{eqnarray}
 && Z_{K3}(\tau,\bar{\tau};z,\bar{z}) 
=  \frac{e^{-4\pi \frac{z_2^2}{\tau_2}}}{2N}\sum_{I,\tI} \sum_{\al}
  N_{I,\tI} F_I^{(\al)}(\tau,z) \overline{F_{\tI}^{(\al)}(\tau,z)}~,
\label{Z K3}
\end{eqnarray}
where the sum of $\al$ runs over the spin structures, and 
the angle variables $z$, $\bar{z}$ couple with
the total $U(1)_R$-charge.
The factor $ e^{- 4 \pi z_2^2/\tau_2}$ ($\tau_2\equiv {\rm Im} \tau$, $z_2\equiv {\rm Im} z$) 
is necessary for preserving the modular invariance for $z\neq 0$; 
it is related to the chiral anomaly of the total $U(1)_R$ current.
The chiral blocks $F^{(\al)}_I(\tau,z)$ are explicitly written 
as `integral spectral flow orbits' \cite{EOTY} as 
\begin{eqnarray}
 && F^{(\sNS)}_I(\tau,z) 
= 
\frac{1}{N}
\sum_{a,b\in\bsz_N} q^{a^2}y^{2a} \prod_{i=1}^r
\ch{(\sNS), k_i}{\ell_i,m_i}(\tau,z+a\tau+b)~, 
\label{F I NS}
\end{eqnarray}
for the NS sector. 
Here $I$ is the collective index:
$I \equiv \{ (\ell_1,m_1), \ldots, (\ell_r,m_r)\}$, 
and likewise for $\tilde I$ ($0\leq \ell_i \leq k_i$, 
$m_i \in \bz_{2(k_i+2)}$, $\ell_i+m_i \in 2\bz$).
The characters $\ch{(\sNS),k_i}{\ell_i,m_i}(\tau,z)$ of the 
$\cN=2$ minimal models $M_{k_i}$ are presented in Appendix A.
The orbits of the other spin structures 
are defined with $1/2$-spectral flows\footnote
  {In the convention taken here, we do not include extra 
   phase factors originating from the $U(1)_R$-charge. 
   Consequently, our chiral blocks of the $\tNS$ and $\tR$ sectors
   have slightly unnatural $q$-expansions for some $I$, such as
$$
 F^{(\stNS)}_I(\tau) = - q^{h_I} + a_1 q^{h_I+1} + 
  a_2 q^{h_I+2} \cdots~.
$$
  Also, the collective index $I\equiv \{ (\ell_1,m_1), \ldots,
   (\ell_r,m_r) \}$ encodes quantum numbers of the NS sector
  even in $F^{(\sR)}_I$ and $F^{(\stR)}_I$.
  An advantage of this convention is that the modular 
  S-matrices are common to all spin structures.
}:
\begin{eqnarray}
 && F^{(\stNS)}_I(\tau,z) = F^{(\sNS)}_I(\tau,z+\frac{1}{2})
~, \nn
&& F^{(\sR)}_I(\tau,z) = q^{\frac{1}{4}}y 
F^{(\sNS)}_I(\tau,z+\frac{\tau}{2}) ~, \nn
&& F^{(\stR)}_I(\tau,z) = q^{\frac{1}{4}}y 
F^{(\sNS)}_I(\tau,z+\frac{\tau}{2}+\frac{1}{2}) ~,
\label{F I}
\end{eqnarray}
The multiplicity $N_{I,\tI}$ is simply\footnote
  {We start with the A-type modular invariant for each minimal model
  $M_{k_i}$ for simplicity.
The second term in \eqn{N I tI} is due to the `field 
identification'
$$
\ch{(\sNS),k_i}{k_i-\ell_i,m_i+k_i+2}(\tau,z)
=\ch{(\sNS),k_i}{\ell_i,m_i}(\tau,z)~.
$$
},
\begin{eqnarray}
 && N_{I,\tI} \equiv \prod_{i=1}^r\, \frac{1}{2} \left(
\delta_{\ell_i,\tilde{\ell}_i} \delta^{(2(k_i+2))}_{m_i,\tilde{m}_i}
+ \delta_{\ell_i,k_i-\tilde{\ell}_i} \delta^{(2(k_i+2))}_{m_i,\tilde{m}_i+k_i+2}
\right) ~.
\label{N I tI} 
\end{eqnarray}
The summation over $b \in \bz_{N}$ in \eqn{F I NS} 
projects out states that do not satisfy the $U(1)$-charge 
integrality condition,
\begin{eqnarray}
&& Q(I) \equiv \sum_{i=1}^r \frac{m_i}{k_i+2} \in \bz~,
\label{U(1) charge constraint} 
\end{eqnarray}
which is necessary for the space-time SUSY.
By construction $F^{(\sNS)}_{I}$ vanishes unless \eqn{U(1) charge constraint}
is satisfied.
On the other hand the integral spectral flow ($a\in\bz_{N}$) acts on the collective 
index $I$ as 
\begin{eqnarray}
 s~:~ I\equiv \{(\ell_1,m_1),\cdots, (\ell_r,m_r)\}~ \longmapsto ~
 s(I)\equiv \{(\ell_1,m_1-2),\cdots, (\ell_r,m_r-2)\}~,
\end{eqnarray}
so obviously,
\begin{eqnarray}
 F^{(\al)}_{s^n(I)}(\tau,z) = F^{(\al)}_{I}(\tau,z) 
~, ~~~ ({}^{\forall}n\in \bz)~.
\end{eqnarray}
In this sense the summation in \eqn{Z K3} overcounts the chiral blocks
and the factor of $1/N$ has been included to compensate the redundancy. 
The chiral blocks $F^{(\sNS)}_{I}(\tau,z)$ defined this way are
often useful,  since the modular invariance is manifest.


To close this preliminary section, we briefly illustrate the structure
of the Hilbert space in the Gepner construction of $K3$.  
By the above construction the Hilbert spaces are shown to be 
\begin{eqnarray}
 && \cH^{(\al)}_{\msc{Gepner}} = \bigoplus_{n\in \bsz_N} 
\bigoplus_{\stackrel{\scriptstyle I,\tI}{Q(I)\in \bz,~ Q(\tI) \in \bz}} \,\left\lb 
N_{I,\tI} \, \cH^{(\al)}_{s^n(I),L} \otimes 
\cH^{(\al)}_{\tI,R} \right\rb ~, ~~~ (\al =\NS, \R)
\label{H Gepner}
\end{eqnarray}
where 
$s^n$ is the actions of the integral spectral flows and
$\cH^{(\al)}_{I,L}$ ($\cH^{(\al)}_{\tI,R}$) denotes the left (right) moving
Hilbert spaces corresponding to the chiral blocks $F^{(\al)}_I(\tau,z)$
($\overline{F^{(\al)}_{\tI}(\tau,z)}$), that are tensor products of the
$M_{k_i}$ minimal model Hilbert spaces.
Note that the left-right symmetric primary states lie in the  
$n=0$ sector, but we also have many asymmetric primary states generated
by the spectral flows. 
We will later work with the type II string vacua that include chiral
spin structures. In those cases the Hilbert spaces \eqn{H Gepner} need
be extended by the $1/2$-spectral flows acting chirally. 

In the present $\hat{c}=2$ case relevant for $K3$,
the $\cN=2$ superconformal symmetry is enhanced to the (small) $\cN=4$ 
by adding the spectral flow operators, which are identified with 
the $SU(2)_1$ currents $J^{\pm} \equiv J^1\pm i J^2$ in the $\cN=4$
superconformal algebra (SCA) \cite{EOTY}. Accordingly, the chiral parts of 
$\cH^{(\al)}_{\msc{Gepner}}$ are decomposed into irreducible
representations of $\cN=4$ SCA at level 1, that are classified as follows 
\cite{ET}:
\begin{itemize}
 \item {{\bf massive representations:} $\cC^{(\sNS)}_h$, $\cC^{(\sR)}_h$}

These are non-degenerate representations whose vacua have   
conformal weights $h$. 
The vacuum of $\cC^{(\sNS)}_h$
belongs to the spin $0$ representation of the $SU(2)_1$-symmetry. 
The four-fold degenerate vacua of $\cC^{(\sR)}_h$ 
generate the representation $2[\mbox{spin 0}] \oplus [\mbox{spin 1/2}]$.
Unitarity requires $h\geq 0$ for $\cC^{(\sNS)}_h$ and $h\geq
\frac{1}{4}$ for $\cC^{(\sR)}_h$.
The 1/2-spectral flow connects $\cC^{(\sNS)}_h$ with $\cC^{(\sR)}_{h+\frac{1}{4}}$.
 \item {{\bf massless representations:} $\cD^{(\sNS)}_{\ell}$, 
$\cD^{(\sR)}_{\ell}$  ($\ell=0,1/2$)}

These are degenerate representations whose vacua have conformal weights 
$h=\ell$ for the NS representations $\cD^{(\sNS)}_{\ell}$, and 
$h=\frac{1}{4}$ for the Ramond representations $\cD^{(\sR)}_{\ell}$;
they belong to the spin $\ell$ representation of $SU(2)_1$.
To be more specific, $\cD^{(\sNS)}_0$
(`graviton rep.' or `identity rep.') corresponds to the unique vacuum
with $h=0$, $J^3_0=0$, while $\cD^{(\sNS)}_{1/2}$ (`massless matter rep.') is
generated over doubly degenerated vacua with $h=1/2$, $J^3_0=\pm 1/2$.
The Ramond sector $\cD^{(\sR)}_{\frac{1}{2}-\ell}$ is connected with $\cD^{(\sNS)}_{\ell}$
by the 1/2-spectral flow.

\end{itemize}
The decomposition in terms of the $\cN=4$ SCA will be crucial for our
construction of the mirrorfolds. 
The relevant character formulas are summarized in Appendix A.


~

\subsection{The Mirror Twist}



Now let us specify the precise action of the mirror involution 
operator $\sigma_{\msc{mirror}}$ in \eqn{mirrorfold}. 


First of all, 
$\sigma_{\msc{mirror}}$ should act as the $U(1)$-charge conjugation
in the right moving $\cN=2$ SCA:
\begin{eqnarray}
 && \sigma_{\msc{mirror},R} (\equiv \sigma^{\cN=2}_R)~:~ 
T_R\,\rightarrow\,T_R~,~~J_R\,\rightarrow \,-J_R~,~~
G_R^{\pm}\,\rightarrow\, G_R^{\mp}~,
\label{sigma R N=2}
\end{eqnarray}
while leaving the left moving $\cN=2$ SCA unchanged. 
Moreover, as the theory is endowed with the $\cN=4$ SCA at level 1, 
the above $\sigma_{\msc{mirror},R}$ acts on the right-moving $\cN=4$ generators 
$\{T_R,G^a_R,J^i_R\}$ ($a=0,1,2,3$ and $i=1,2,3$) as well.
With the generators of the (total) $\cN=2$ SCA identified as 
\begin{eqnarray}
&&
J_R = 2J_R^3~, ~~~ G^{\pm}_R = G^{0}_R\pm i G^3_R~,
\end{eqnarray}
the action of the involution is naturally extended on the $\cN =4$ algebra as
\begin{eqnarray}
\sigma_{\msc{mirror}, R} (\equiv \sigma^{\cN=4}_{1,R})
~:~ && T_R\,\rightarrow\,T_R~, ~~ 
J_R^1\,\rightarrow \, J_R^1~, ~~ 
J_R^i\,\rightarrow \, -J_R^i ~(i=2,3)~, ~~ \nn
&& 
 G_R^a\,\rightarrow\, G_R^a ~ (a=0,1)~, ~~
 G_R^a\,\rightarrow\, -G_R^a ~ (a=2,3)~.
\label{sigma R N=4}
\end{eqnarray}
Here we have introduced the symbol 
$\sigma^{\cN=4}_{1,R}$ for later convenience, 
and $\sigma^{\cN=4}_{2,R}$, $\sigma^{\cN=4}_{3,R}$ are defined in the same way 
by the cyclic permutations of the indices $i$ and $a$. 
Since we are assuming the Gepner construction, 
the total involution $\sigma_{\msc{mirror},R}$
is most naturally realised by taking the tensor product of $\cN=2$ involutions in
each $\cN=2$ minimal model $M_{k_i}$ ($i=1,\ldots, r$):
\begin{eqnarray}
 && \sigma_{\msc{mirror},R} \equiv  \prod_{i=1}^r \sigma^{\cN=2, (i)}_R ~,
\label{sigma mirror R}
\end{eqnarray}
where $\sigma^{\cN=2, (i)}_R$ acts on the $\cN=2$ SCA of $M_{k_i}$ as 
\begin{eqnarray}
 && \sigma^{\cN=2, (i)}_{R}~:~T^{(i)}_R \, \rightarrow \, 
T^{(i)}_R~, ~~
J^{(i)}_R\,\rightarrow\, -J^{(i)}_R~,~~ G^{\pm,(i)}_R\, 
\rightarrow \, 
G^{\mp,(i)}_R ~.
\label{N=2 sigma}
\end{eqnarray}
It is easy to see that $\sigma_{\msc{mirror},R}$ defined in this way  
acts on the $\cN=4$ SCA as the operator $\sigma^{\cN=4}_{1,R}$ above. 
We shall assume the right-moving operation of the form 
\eqn{sigma mirror R} from now on.

The operation of the left-mover $\sigma_{\msc{mirror}, L}$ still needs to be determined.
The simplest guess would be $\sigma_{\msc{mirror}, L} \equiv {\bf 1}$,
but this does not work.
In fact, it turns out that 
$\sigma^{\msc{naive}}_{\msc{mirror}}\equiv {\bf 1} \otimes 
\sigma_{\msc{mirror}, R}$ 
does not leave invariant the closed string Hilbert space of the Gepner model 
$\cH_{\msc{Gepner}}$.\footnote
  {For example, pick up a symmetric primary state of the form
$$
\ket{v} \equiv \prod_{i}\ket{\ell_i,m_i,s_i}_L \otimes 
\prod_{i}\ket{\ell_i,m_i,s_i}_R~, ~~~ 
(\ell_i+m_i+s_i \in 2 \bz,~ m_i \in \bz_{2(k_i+2)}, ~ s_i \in \bz_4).
$$
The above $\sigma^{\msc{naive}}_{\msc{mirror}}$ acts on it as 
$$
\sigma^{\msc{naive}}_{\msc{mirror}} \ket{v} = 
\prod_{i}\ket{\ell_i,m_i,s_i}_L \otimes 
\prod_{i}\ket{\ell_i,- m_i,-s_i}_R~,
$$
which in general is not a state in
$\cH_{\msc{Gepner}}$. 
}
We propose that the operator $\sigma_{\msc{mirror},L}$ should satisfy
following requirements:
\begin{enumerate}
 \item $\sigma_{\msc{mirror}}\equiv \sigma_{\msc{mirror},L} \otimes 
\sigma_{\msc{mirror},R}$ acts over $\cH_{\msc{Gepner}}$ as an involution, 
\begin{eqnarray}
 \sigma_{\msc{mirror}} (\cH_{\msc{Gepner}}) = \cH_{\msc{Gepner}}, ~~~
 \left(\sigma_{\msc{mirror}}\right)^2 = {\bf 1}.
\end{eqnarray}
 \item $\sigma_{\msc{mirror},L}$ preserves the total 
$\cN=2$ SCA $\{T_L,J_L,G^{\pm}_L\}$. 
 \item The orbifolding by 
$\sigma_{\msc{mirror}}\equiv \sigma_{\msc{mirror},L}\otimes
 \sigma_{\msc{mirror},R} $ is compatible with modular 
invariance. 
\end{enumerate}
Due to the second requirement, $\sigma_{\msc{mirror},L}$ can only act as a 
linear transformation on the primary states 
of the total $\cN=2$ SCA. 
Especially, it can be regarded as phase changes on a suitably 
chosen basis of primary states.

We consider following two candidates for $\sigma_{\msc{mirror},L}$:

\begin{description}
\item [(i)]
$\sigma_{\msc{mirror},L}$ acting on the $\cN=4$ SCA as the
automorphism $\sigma^{\cN=4}_{3,L}$. 
For the $\cN=4$ primary states $\ket{v}_L$, the action of 
$\sigma_{\msc{mirror},L}$ is defined as
\begin{equation}
\sigma_{\msc{mirror},L} \ket{v}_L \equiv \left\{
\begin{array}{ll}
\prod_{i=1}^r \sigma^{\cN=2, (i)}_{L} \ket{v}_L~, &
(2J^3_{L,0}\ket{v}_L=0)~, \\
J^+_{L,0} \prod_{i=1}^r \sigma^{\cN=2, (i)}_{L} \ket{v}_L~, &
(2J^3_{L,0}\ket{v}_L=\ket{v}_L)~, \\
- J^-_{L,0} \prod_{i=1}^r \sigma^{\cN=2, (i)}_{L} \ket{v}_L~, &
(2J^3_{L,0}\ket{v}_L=-\ket{v}_L)~,
\end{array}\right.
\label{sigma primary L 1}
\end{equation}
where $J^{\pm}_L \equiv J^1_L \pm i J^2_L$ are the $SU(2)$
currents in the $\cN=4$ SCA.


\item [(ii)]
$\sigma_{\msc{mirror},L}$  preserving the $\cN=4$ SCA.
For the $\cN=4$ primary states $\ket{v}_L$, the action of 
$\sigma_{\msc{mirror},L}$ is defined as
\begin{equation}
\sigma_{\msc{mirror},L} \ket{v}_L \equiv \left\{
\begin{array}{ll}
\prod_{i=1}^r \sigma^{\cN=2, (i)}_{L} \ket{v}_L~, &
(2J^3_{L,0}\ket{v}_L=0)~, \\
J^+_{L,0} \prod_{i=1}^r \sigma^{\cN=2, (i)}_{L} \ket{v}_L~, &
(2J^3_{L,0}\ket{v}_L=\ket{v}_L)~, \\
J^-_{L,0} \prod_{i=1}^r \sigma^{\cN=2, (i)}_{L} \ket{v}_L~, &
(2J^3_{L,0}\ket{v}_L=-\ket{v}_L)~.
\end{array}\right.
\label{sigma primary L 2}
\end{equation}
\end{description}
It is easy to verify that these two candidates indeed satisfy 
the first and second conditions given above. 
Checking the modular invariance is a non-trivial task and we will discuss it from now on.


~


\subsection{Modular Invariant Partition Functions of the Mirrorfolds with
  Compact $K3$ Fibers}

We shall construct modular invariant partition functions of the mirrorfolds \eqn{mirrorfold}. 
We take an arbitrary Gepner model describing a compact $K3$ fiber. 
We assume \eqn{sigma mirror R}
for $\sigma_{\msc{mirror}, R}$, and adopt the first candidate \eqn{sigma
primary L 1} for $\sigma_{\msc{mirror}, L}$.


Before discussing the construction of the modular invariant, 
we need to find the $\cN=4$ character formulas twisted
by $\sigma^{\cN=4}_{i,L}$ ($\sigma^{\cN=4}_{i,R}$). 
We first focus on the $\sigma^{\cN=4}_{3,L}$-twist. 
%
%
We express the spatial and temporal boundary conditions 
as $[S,T]$, $S,T \in \bz_2$
($S,T=0$ means no twist, while $S,T=1$ indicates 
twisting by $\sigma^{\cN=4}_{3,L}$).
The desired character formulas are readily obtained by starting with the temporal 
twist boundary condition $[S,T]=[0,1]$ (i.e. inserting $\sigma^{\cN=4}_{3,L}$ into 
the trace), which results in an extra phase factor $(-1)^n$ in the $n$-th 
spectral flow sector.  
For $[S,T]=[0,1]$ the formula \eqn{decomp N=4 ch} is thus replaced by
\begin{eqnarray}
\ch{\cN=4,(\sNS)}{*,[0,1]}(*;\tau,z) 
&\equiv& \tr_{\cH}\lb\sigma^{\cN=4}_{3,L}q^{L_0-\frac{1}{4}} y^{2J^3_0}\rb\nn
&=& \sum_{n\in \bz} (-1)^n q^{n^2}y^{2n}
\ch{\cN=2,(\sNS)}{*}(*;\tau,z+n\tau)~.
\label{twisted N=4 ch 0}
\end{eqnarray}
Here $\cH$ denotes the representation space of $\cC^{(\sNS)}_h$, 
$\cD^{(\sNS)}_0$ or
$\cD^{(\sNS)}_{1/2}$. 
We spell out explicit results in each case:
\begin{description}
 \item[massive representation :]
\begin{eqnarray}
\ch{\cN=4,(\sNS)}{[0,1]}(h;\tau,z) &= &q^{h-\frac{1}{8}} \sum_{n\in\bz}
 (-1)^n q^{\frac{n^2}{2}}y^n \frac{\th_3(\tau,z)}{\eta(\tau)^3}
= q^{h-\frac{1}{8}}
\frac{\th_3(\tau,z)\th_4(\tau,z)}{\eta(\tau)^3}~,
\label{twisted N=4 massive} 
\end{eqnarray} 
\item[massless representations :]
\begin{eqnarray} 
 \ch{\cN=4,(\sNS)}{0, [0,1]}(\ell=\frac 12;\tau,z) &= &
q^{-1/8}\,
\sum_{n\in \bsz}\, (-1)^{n+1}\frac{1}{1+yq^{n-1/2}}\, 
q^{\frac{n^2}{2}}y^n
\frac{\th_3(\tau,z)}{\eta(\tau)^3}~,
\label{twisted N=4 massless} \\
\ch{\cN=4,(\sNS)}{0, [0,1]}(\ell=0;\tau,z) &= &
q^{-1/8}\,
\sum_{n\in \bsz}\,  (-1)^n
 \frac{(1-q)q^{\frac{n^2}{2}+n-\frac{1}{2}}y^{n+1}}
{(1+yq^{n+1/2})(1+yq^{n-1/2})} 
\frac{\th_3(\tau,z)}{\eta(\tau)^3} \nn
&\equiv& q^{-1/8} 
\frac{\th_3(\tau,z)\th_4(\tau,z)}{\eta(\tau)^3}
\equiv \ch{\cN=4,(\sNS)}{[0,1]}(h=0;\tau,z)
~.
\label{twisted N=4 grav} 
\end{eqnarray}
The second line of \eqn{twisted N=4 grav} follows from identity
\begin{eqnarray}
 && \frac{(1-q)q^{n-\frac{1}{2}}y}
{(1+yq^{n+1/2})(1+yq^{n-1/2})} 
= 1- \frac{1}{1+yq^{n-\frac{1}{2}}} -
\frac{yq^{n+\frac{1}{2}}}{1+yq^{n+\frac{1}{2}}}~.
\label{identity 1}
\end{eqnarray}
\end{description}
Specializing to $z=0$, we further obtain 
\begin{eqnarray}
 && \hspace{-1cm}
\ch{\cN=4,(\sNS)}{[0,1]}(h;\tau,0)
= q^{h-1/8}\frac{\th_3(\tau)\th_4(\tau)}{\eta(\tau)^3} \equiv
 q^{h-1/8} \frac{2}{\th_2(\tau)} 
\equiv \chi_{[0,1]}(p;\tau)~, 
~~~ (h= \frac{p^2}{2}+\frac{1}{8}) ~, 
\label{twisted N=4 massive 2} \\
 &&  \hspace{-1cm}
\ch{\cN=4,(\sNS)}{0, [0,1]}(\ell=1/2;\tau,0) = 
q^{-1/8}\,
\sum_{n\in \bsz}\, (-1)^{n+1}\frac{1}{1+q^{n-1/2}}\, 
q^{\frac{n^2}{2}}
\frac{\th_3(\tau)}{\eta(\tau)^3} \equiv 0~,
~~~ (h=1/2)~,
\label{twisted N=4 massless 2}\\
 && \hspace{-1cm}
\ch{\cN=4,(\sNS)}{0,[0,1]}(\ell=0;\tau,0)
= q^{-1/8}\frac{\th_3(\tau)\th_4(\tau)}{\eta(\tau)^3} \equiv
 q^{-1/8} \frac{2}{\th_2(\tau)} \equiv 
\chi_{[0,1]}(p=i/2;\tau)~,  (h=0)~,
\label{twisted N=4 grav 2} 
\end{eqnarray}
where we used the abbreviation $\th_i(\tau)\equiv \th_i(\tau,0)$, and 
$\chi_{[0,1]}(p;\tau)$ is the $\cN=2$ twisted character of $\hat{c}=2$
\eqn{twisted massive}.


The character formulas for the other boundary conditions are 
derived by acting modular transformations successively, at least for $z=0$.
We denote the spin structures and the boundary conditions 
of $\sigma^{\cN=4}_{3,L}$ such as $\{\NS,\, [S,T]\}$. 
Starting with the character formula of $\{\NS,\,[0,1]\}$ given above, 
it turns out that there are three types of non-trivial characters 
$\chi_{[0,1]}(p;\tau)$, $\chi_{[1,0]}(p;\tau)$, $\chi_{[1,1]}(p;\tau)$
(see \eqn{twisted massive}):
\begin{eqnarray}
&&  \{\NS,\, [0,1]\}, ~  \{\tNS,\, [0,1]\}~:~~  
\chi_{[0,1]}(p;\tau) \equiv \frac{2 q^{\frac{p^2}{2}}}{\th_2(\tau)}~, ~~~ 
(h=\frac{p^2}{2}+\frac{1}{8})~, \nn
&&
 \{\NS,\, [1,0]\}, ~  \{\R,\, [1,0]\} ~:~~
\chi_{[1,0]}(p;\tau) \equiv \frac{2 q^{\frac{p^2}{2}}}{\th_4(\tau)}~, ~~~ 
(h=\frac{p^2}{2}+\frac{1}{4})~, \nn
&&
\{\tNS,\, [1,1]\}, ~  \{\R,\, [1,1]\} ~:~~  
\chi_{[1,1]}(p;\tau) \equiv \frac{2 q^{\frac{p^2}{2}}}{\th_3(\tau)}~, ~~~ 
(h=\frac{p^2}{2}+\frac{1}{4})~.
\label{twisted N=4 characters}
\end{eqnarray}
These are the building blocks necessary for our construction of the 
mirrorfold modular invariants. 
There still remain boundary conditions that are connected  
to $\{\R,\, [0,1]\}$ and  $\{\tR,\, [0,1]\}$ by modular transformations. 
We need some further technicality to obtain such twisted characters, 
and the complete list of the $\cN=4$ twisted characters are given in Appendix D. 
For our purposes, however, only the ones given in (\ref{twisted N=4 characters}) are needed.


What about the $\sigma^{\cN=4}_{1,L}$-twisting? 
Since the $\sigma^{\cN=4}_{1,L}$-twist acts as $J(\equiv 2J^3)\,\rightarrow\, -J$
on the $U(1)_R$-current of the underlying $\cN=2$ SCA, none of the
spectrally flowed sectors contribute 
to the $\sigma^{\cN=4}_{1,L}$-twisted characters. 
Recalling that the $\cN=4$ SCA is obtained by extending the $\cN=2$ SCA by adding 
the spectral flow operators, 
we conclude that the $\sigma^{\cN=4}_{1,L}$-twisted $\cN=4$
characters must coincide with the twisted $\cN=2$ characters of
$\hat{c}=2$ \eqn{twisted massive}. 
This means that we are simply led to the same classification of 
$\sigma^{\cN=4}_1$-twisted characters as \eqn{twisted N=4 characters}.

In this sense it seems natural to express the above twisted character
$\chi_{[0,1]}(p;\tau)$ in two different ways,
one that is natural for the $\sigma^{\cN=4}_3$-twist, and the other 
for the $\sigma^{\cN=4}_1$ twist:
\begin{eqnarray}
 \chi_{[0,1]}(p;\tau) 
&=& \tr_{\cC^{(\sNS)}_h}\lb \sigma^{\cN=4}_{1,L}q^{L_0-\frac{1}{4}} \rb
= \frac{q^{h-1/8}}{\eta(\tau)} \cdot
\sqrt{\frac{2\eta(\tau)}{\th_2(\tau)}}\cdot
\sqrt{\frac{\th_3(\tau)\th_4(\tau)}{\eta(\tau)^2}} 
\nn
&\equiv & q^{h-\frac{1}{4}} 
\frac{\prod_{n=1}^{\infty}(1+q^{n-1/2})(1-q^{n-1/2})}
{\prod_{n=1}^{\infty}(1-q^n)(1+q^n)} ~,
\label{chi p 1} \\
 \chi_{[0,1]}(p;\tau) 
&=& \tr_{\cC_h^{(\sNS)}}\lb \sigma^{\cN=4}_{3,L}q^{L_0-\frac{1}{4}} \rb
= q^{h-1/8} \th_4(\tau) \cdot \frac{\th_3(\tau)}{\eta(\tau)^3}
\nn
&\equiv & q^{h-\frac{1}{4}}
\sum_{n\in \bz}(-1)^n q^{\frac{n^2}{2}}\,
\frac{\prod_{n=1}^{\infty}(1+q^{n-1/2})^2}
{\prod_{n=1}^{\infty}(1-q^n)^2} ~.
\label{chi p 3}
\end{eqnarray}
The equality of \eqn{chi p 1} and \eqn{chi p 3} is immediately checked by
the Euler identity $\th_2(\tau)\th_3(\tau)\th_4(\tau)= 2\eta(\tau)^3$.
%
The equivalence of the $\sigma^{\cN=4}_3$- and $\sigma^{\cN=4}_1$-twisted character formulas
\eqn{twisted N=4 characters} is anticipated from the existence of an automorphism 
interpolating $\sigma^{\cN=4}_3$ and $\sigma^{\cN=4}_1$ within the $\cN=4$ SCA.
Similar results for the other boundary condition (such as $\{\R,\,[0,1]\}$), which are less trivial, are discussed in Appendix D. 


~

We now proceed to our main analysis.  
The chiral blocks for each sector of the $K3$ twisted by 
$\sigma_{\msc{mirror}}\equiv \sigma_{\msc{mirror},L}\otimes 
\sigma_{\msc{mirror},R}$ are obtained as follows.

~

\noindent
{\bf The right-mover }

After making the $\sigma_{\msc{mirror},R}$-insertion 
only the spectral flow orbits of type 
$\{(\ell_1,0),\ldots, (\ell_r,0)\}$ belonging to the $\NS$ or $\tNS$ sectors
survive, 
while none of the orbits in the $\R$ nor $\tR$ sectors
contributes\footnote
   {The easiest way to see this is to recall that the $\cN=2$ involution
     $\sigma^{\cN=2}$ acts on primary states of the $\cN=2$ minimal
     model $M_k$ as 
$$
\sigma^{\cN=2}~:~ \ket{\ell,m,s}~\longmapsto ~ \ket{\ell,-m,-s}~, ~~~
  (\ell+m+s \in 2\bz, ~~~ \ell=0,\ldots, k, ~ m\in \bz_{2(k+2)},~ s\in \bz_4)~.
$$
In the R-sector we have $s=\pm 1 ~(\mod\, 4)$ and thus 
$\sigma^{\cN=2}$ does not have any fixed point. 
This means that the $\sigma^{\cN=2}$-inserted traces always vanish in the R-sector. 
}. 
The resultant chiral blocks are 
\begin{eqnarray}
&& \overline{\chi^{\bk}_{\bl, [S,T]} (\tau)}  \equiv \prod_{i=1}^r 
\overline{\chi^{k_i}_{\ell_i,[S,T]}(\tau)}~, 
\label{chi bk bl}
\\
&& \bl \equiv (\ell_1,\ldots, \ell_r)~, ~~~
\bk \equiv (k_1,\ldots, k_r)~, \nonumber
\end{eqnarray}
where $\chi^{k_i}_{\ell_i,[S,T]}(\tau)$ are the twisted characters of 
the $\cN=2$ minimal models \eqn{twisted minimal}.
The chiral blocks \eqn{chi bk bl} can also be expressed 
in terms of the twisted $\cN=4$ characters $\chi_{[S,T]}(p;\tau)$ 
\eqn{twisted N=4 characters}.
For example, picking up the spectral flow orbit
$\bl \equiv \{ (\ell_1,0), \ldots, (\ell_r,0)\}$, we may write,
\begin{eqnarray}
 \overline{\chi^{\bk}_{\bl,[0,1]}(\tau)} &\equiv&
\tr_{\msc{orbit of $\bl$}}\left\lb 
\sigma_{\msc{mirror},R} \, \bar{q}^{\tL_0-\frac{1}{4}}
\right\rb  \nn
&=& \sum_{n=0}^{\infty} a_{n,\bl} \, \overline{\chi_{[0,1]}(p_{n,\bl};\tau)}
\hspace{1cm}
\left(
\frac{p_{n,\bl}^2}{2}+\frac{1}{8} = h_{\bl}+n ~, ~~
 h_{\bl} \equiv \sum_{i} \frac{\ell_i(\ell_i+2)}{4(k_i+2)}
\right) \nn
& =&  \sum_{n=0}^{\infty} a_{n,\bl} \,
\bar{q}^{h_{\bl}+n-\frac{1}{8}}
\frac{2}{\overline{\th_2(\tau)}}~,
\end{eqnarray}
with $a_{n,\bl} \in \bz$, $a_{0,\bl}=1$. 
It is convenient to introduce a function $f^{\bk}_{\bl,[0,1]}(\tau)$
defined by power series
\begin{eqnarray}
 && f^{\bk}_{\bl,[0,1]}(\tau) \equiv \sum_{n=0}^{\infty} 
a_{n,\bl} \, q^{h_{\bl}+n-\frac{1}{8}}~, 
\end{eqnarray}
or more concisely, 
\begin{eqnarray}
 && 
\chi^{\bk}_{\bl,[0,1]}(\tau)
= \frac{2}{\th_2(\tau)} f^{\bk}_{\bl,[0,1]}(\tau)~. 
\label{f l k}
\end{eqnarray}
Similar functions for the other boundary conditions
$f^{\bk}_{\bl,[1,0]}$, $f^{\bk}_{\bl,[1,1]}$
are defined in the same way,
\begin{eqnarray}
&& \chi^{\bk}_{\bl,[1,0]}(\tau)
= \frac{2}{\th_4(\tau)} f^{\bk}_{\bl,[1,0]}(\tau)~,
~~~
\chi^{\bk}_{\bl,[1,1]}(\tau)
= \frac{2}{\th_3(\tau)} f^{\bk}_{\bl,[1,1]}(\tau)~.
\label{f l k 2}
\end{eqnarray}
By construction we find,
\begin{eqnarray}
&& \tr_{\cN=4 ~\msc{vacua of}~  \bl} \lb \sigma_{\msc{mirror},R} \, 
\bar{q}^{\tL_0-\frac{1}{4}} \rb = 
\overline{f^{\bk}_{\bl,[0,1]}(\tau)} ~,
\label{trace fkl}
\end{eqnarray}
where the trace is taken over the $\cN=4$ primary states 
belonging to the orbit $\bl$. 
Modular properties of functions $f^{\bk}_{\bl,[S,T]}(\tau)$ are 
immediately read off from those of the $\cN=2$ twisted minimal characters 
$\chi^{k_i}_{\ell_i,[S,T]}$. See formulas \eqn{modular twisted minimal}.

~


\noindent
{\bf The left-mover }

Since we have assumed \eqn{sigma primary L 1} for $\sigma_{\msc{mirror},L}$, 
it is convenient to decompose the chiral blocks into the $\cN=4$
irreducible representations. 
Contributions from the massless rep. $\cD^{(\sNS)}_{1/2}$
$(Q= \pm 1)$ trivially vanish because of \eqn{twisted N=4 massless 2}. 
Also, the Ramond rep. $\cD^{(\sR)}_{1/2}$ does not contribute because 
\begin{eqnarray}
 && \ch{\cN=4,(\sR)}{0,[0,1]}(\ell=1/2;\tau,0)
\left(\equiv \tr_{\cD^{(\sR)}_{1/2}}\left\lb 
\sigma^{\cN=4}_{3,L} q^{L_0-\frac{1}{4}}
\right\rb \right) \equiv  q^{\frac{1}{4}} \ch{\cN=4,(\sNS)}{0,[0,1]}
(\ell=0;\tau, \frac{\tau}{2}) \nn
&& \hspace{4cm} = q^{-\frac{1}{8}} \frac{i\th_1(\tau,0)\th_2(\tau,0)}
{\eta(\tau)^3} =0~, 
\end{eqnarray}
where we have used \eqn{twisted N=4 grav} in the second line. 
Thus, possible non-vanishing contributions only come from 
representations generated by neutral $\cN=4$ primary states $(Q=0)$.
Since  $\sigma_{\msc{mirror},L}$ acts as 
$\prod_i \sigma^{\cN=2,(i)}_L$ on neutral $\cN=4$ primaries, 
again we find only the contributions from spectral flow orbits 
$\bl \equiv \{(\ell_1,0), \ldots, (\ell_r,0)\}$ in the $\NS$ ($\tNS$)
sector, and no contribution from the $\R$ ($\tR$) sector.  
We thus obtain, 
\begin{eqnarray}
 && 
\tr_{\cN=4 ~\msc{vacua of}~  \bl} \lb \sigma_{\msc{mirror},L} \, 
q^{L_0-\frac{1}{4}} \rb = 
f^{\bk}_{\bl,[0,1]}(\tau) ~, \\
&&
\tr_{\cN=4 ~\msc{vacua of other neutral orbits}} \lb \sigma_{\msc{mirror},L}\, 
q^{L_0-\frac{1}{4}} \rb = 0~. 
\end{eqnarray}
As we observed above, the $\sigma^{\cN=4}_3$-twisted characters 
are equal to the $\sigma^{\cN=4}_1$-twisted ones in the relevant 
sectors. Therefore, we conclude that the chiral blocks of the left-mover formally 
take the same form as the right-mover:
\begin{eqnarray}
 \tr_{\msc{orbit of}~\bl} \lb 
\sigma_{\msc{mirror},L} q^{L_0-\frac{1}{4}} \rb = f^{\bk}_{\bl,[0,1]}(\tau)
\cdot \frac{2}{\th_2(\tau)}
\, \left(\equiv 
f^{\bk}_{\bl,[0,1]}(\tau) \cdot \frac{\th_3(\tau)\th_4(\tau)}{\eta(\tau)^3}
\right) ~. 
\end{eqnarray}
The same happens for other boundary conditions $[1,0]$, $[1,1]$ due to 
modular transformations. 
This fact makes the modular invariance of the $K3$ mirrorfolds possible.

~


At this stage, we may describe the modular 
invariant partition functions for the string vacua 
of our mirrorfold model \eqn{mirrorfold} ($\times$ flat space-time $\br^{4,1}$).
It can be written in the form,
\begin{eqnarray}
 && Z(\tau,\bar{\tau}) = Z^{\msc{u}}(\tau,\bar{\tau})+
Z^{\msc{t}}(\tau,\bar{\tau})~,
\label{Z mirrorfold 0}
\end{eqnarray}
where $Z^{\msc{u}}$ is the partition function of the untwisted sector, 
and $Z^{\msc{t}}$ denotes contributions of the twisted sectors\footnote
{In the literature it is traditional to 
use this term for sectors with only the spatial twist(s). 
Here, we define $Z^{\msc{t}}$ to include also the temporal-twisted sector.
This somewhat non-standard usage is for computational convenience and
hopefully no confusion arises.}
including both temporal and 
spatial twists by $\sigma_{\msc{mirror}}\otimes \cT_{2\pi R}$.

Assuming the type II string vacuum,
the partition function for the untwisted sector is given as 
\begin{eqnarray}
 &&
\hspace{-1cm} 
Z^{\msc{u}}(\tau,\bar{\tau}) = 
\frac{1}{2}\cdot \frac{1}{4N}\sum_{\al,\tal}\sum_{I,\tI} 
 \ep(\al)\ep_{A~\msc{or}~B}(\tal) 
\left(\frac{\th_{\lb \al \rb}}{\eta}\right)^2
\overline{
\left(\frac{\th_{\lb \tal \rb}}{\eta}\right)^2
}
N_{I,\tI} F^{(\al)}_I(\tau)\overline{F^{(\tal)}_{\tI}(\tau)} \cdot
\frac{1}{\tau_2^{3/2}\left|\eta\right|^6} 
Z_{2R}(\tau,\bar{\tau})~, \nn
&&
\label{Z untwisted}
\end{eqnarray}
where we set $\th_{\lb \sNS \rb}=\th_3$, 
$\th_{\lb \stNS \rb}=\th_4$, $\th_{\lb \sR \rb}=\th_2$
($\th_{\lb \stR \rb}=i\th_1\equiv 0$), and 
$\ep(\NS)=\ep(\tR)=+1$, $\ep(\tNS)=\ep(\R)=-1$.
For the right-mover, we set 
$\ep_B(\tal)=\ep(\tal)$ for type IIB, while $\ep_A(\NS)=+1$,
$\ep_A(\tNS)=\ep_A(\R)=\ep_A(\tR)=-1$ for type IIA.
We used abbreviation $F^{(\al)}_I(\tau)\equiv 
F^{(\al)}_I(\tau,0)$ here.

Free non-compact bosons in the (transverse part of) $\br^{4,1}$ 
contribute to the factor $1/\tau_2^{3/2}\left|\eta\right|^6$. 
The familiar partition function of a compact boson with radius 
$R$ is
\begin{eqnarray}
 && Z_R(\tau,\bar{\tau}) = 
\frac{R}{\sqrt{\tau_2}\left|\eta(\tau)\right|^2}
\sum_{w,m\in\bsz} e^{-\frac{\pi R^2}{\tau_2}
\left|w\tau+m\right|^2}~.
\label{Z R}
\end{eqnarray}
We further introduce
\begin{eqnarray}
 && Z_{R,(a,b)}(\tau,\bar{\tau}) = 
\frac{R}{\sqrt{\tau_2}\left|\eta(\tau)\right|^2}
e^{-\frac{\pi R^2}{\tau_2}\left|a\tau+b\right|^2}~, ~~~
(a,b \in \bz)~,
\label{Z R a b}
\end{eqnarray}
which describes the contribution from each winding sector
\begin{eqnarray}
 && Y(z+1,\bar{z}+1) = Y(z,\bar{z}) + 2\pi a R~, \nn
 && Y(z+\tau,\bar{z}+\bar{\tau}) = Y(z,\bar{z}) + 2\pi b R~.
\end{eqnarray}
The sectors with even windings $a,b\in 2\bz$ are identified with 
the untwisted sectors, leading to
\begin{eqnarray}
 \sum_{a,b\in 2\bsz} Z_{R,(a,b)}(\tau,\bar{\tau}) 
= \frac{1}{2}Z_{2R}(\tau,\bar{\tau})~.
\end{eqnarray}


The partition function of the twisted sectors is much more complicated. 
Requiring modular invariance, the partition function $Z^{\msc{t}}(\tau,\bar{\tau})$ is 
expected to be of the form, 
\begin{eqnarray}
 && Z^{\msc{t}}(\tau,\bar{\tau}) 
 = \sum_{\stackrel{\scriptstyle a \in 2\bsz+1}
 {\msc{or}~b\in 2\bsz+1}}\, Z_{R,(a,b)}(\tau,\bar{\tau})\,
 \Xi_{(a,b)}(\tau,\bar{\tau})~,
\end{eqnarray}
where $\Xi_{(a,b)}(\tau,\bar{\tau})$ are some functions that behave
covariantly under modular transformations,
\begin{eqnarray}
 && \Xi_{(a,b)}(\tau+1,\bar{\tau}+1)= \Xi_{(a,b+a)}(\tau,\bar{\tau})~,
  ~~~
\Xi_{(a,b)}(-1/\tau,-1/\bar{\tau})=
\Xi_{(b,-a)}(\tau,\bar{\tau})~.
\label{Xi covariance}
\end{eqnarray}


The winding dependence of $\Xi_{(a,b)}(\tau,\bar{\tau})$ 
primarily originates from the $\sigma_{\msc{mirror}}$-twisting in the 
$K3$-sector:
\begin{description}
 \item[(i)] $a\in 2\bz$, $b\in 2\bz+1$~ :~  the sector with temporal
	    twisting by $\sigma_{\msc{mirror}}$.
 \item[(ii)] $a\in 2\bz+1$, $b\in 2\bz$~ :~  the sector with spatial
	    twisting by $\sigma_{\msc{mirror}}$.
 \item[(iii)] $a\in 2\bz+1$, $b\in 2\bz+1$~ :~  the sector with both
             temporal and spatial twisting by $\sigma_{\msc{mirror}}$.
\end{description}
The calculation for each chiral block of the $K3$ sector is carried
out based on the above argument. 
After summing over the chiral spin structures, we find following 
partition functions for the twisted sectors:
\begin{eqnarray}
&& 
Z^{\msc{t}}(\tau,\bar{\tau}) = \frac{1}{4}
\sum_{\stackrel{\scriptstyle a\in 2\bsz+1}{\msc{or}~b\in 2\bsz+1}}
Z_{R,(a,b)}(\tau,\bar{\tau}) 
\frac{1}{\tau_2^{3/2}\left|\eta\right|^6} 
\, Z^f_{(a,b)}(\tau,\bar{\tau}) \,
\sum_{\bl,\tbl} N^{[[a],[b]]}_{\bl,\tbl}
\chi^{\bk}_{\bl,\,[[a],[b]]}(\tau)
\overline{\chi^{\bk}_{\tbl,\,[[a],[b]]}(\tau)}~\nn
&& 
\hspace{1cm} \equiv  \frac{1}{4}
\sum_{\stackrel{\scriptstyle a\in 2\bsz}{b\in 2\bsz+1}}
Z_{R,(a,b)}(\tau,\bar{\tau}) 
\frac{1}{\tau_2^{3/2}\left|\eta\right|^6} 
\left|
\left(\frac{\th_3}{\eta}\right)^2- 
(-1)^{\frac{a}{2}}\left(\frac{\th_4}{\eta}\right)^2
\right|^2 \nn
&& \hspace{2.5cm}
\times
\sum_{\bl,\tbl} N^{[0,1]}_{\bl,\tbl}
f^{\bk}_{\bl,\,[0,1]}(\tau) 
\th_4(\tau){\frac{\th_3(\tau)}{\eta(\tau)^3}} 
\cdot
\overline{f^{\bk}_{\tbl,\,[0,1]}(\tau)
\frac{1}{\eta(\tau)}
\sqrt{\frac{2\eta(\tau)}{\th_2(\tau)}}
\sqrt{\frac{\th_3(\tau)\th_4(\tau)}{\eta(\tau)^2}}
}
~
\nn
&&
\hspace{1cm}
+ \frac{1}{4}
\sum_{\stackrel{\scriptstyle a\in 2\bsz+1}{b\in 2\bsz}}
Z_{R,(a,b)}(\tau,\bar{\tau}) 
\frac{1}{\tau_2^{3/2}\left|\eta\right|^6} 
\left|
\left(\frac{\th_3}{\eta}\right)^2- 
(-1)^{\frac{b}{2}}\left(\frac{\th_2}{\eta}\right)^2
\right|^2
\nn
&& \hspace{2.5cm}
\times 
\sum_{\bl,\tbl} N^{[1,0]}_{\bl,\tbl}
f^{\bk}_{\bl,\,[1,0]}(\tau) 
\th_2(\tau){\frac{\th_3(\tau)}{\eta(\tau)^3}} 
\cdot
\overline{f^{\bk}_{\tbl,\,[1,0]}(\tau)
\frac{1}{\eta(\tau)}
\sqrt{\frac{2\eta(\tau)}{\th_4(\tau)}}
\sqrt{\frac{\th_2(\tau)\th_3(\tau)}{\eta(\tau)^2}}
}
\nn
&&
\hspace{1cm}
+ \frac{1}{4}
\sum_{\stackrel{\scriptstyle a\in 2\bsz+1}{b\in 2\bsz+1}}
Z_{R,(a,b)}(\tau,\bar{\tau}) 
\frac{1}{\tau_2^{3/2}\left|\eta\right|^3} 
\left|
\left(\frac{\th_4}{\eta}\right)^2+ 
i(-1)^{\frac{a+b}{2}}\left(\frac{\th_2}{\eta}\right)^2
\right|^2 \nn
&& \hspace{2.5cm}
\times
\sum_{\bl,\tbl} N^{[1,1]}_{\bl,\tbl}
f^{\bk}_{\bl,\,[1,1]}(\tau) 
\th_2(\tau){\frac{\th_4(\tau)}{\eta(\tau)^3}} 
\cdot
\overline{f^{\bk}_{\tbl,\,[1,1]}(\tau)
\frac{1}{\eta(\tau)}
\sqrt{\frac{2\eta(\tau)}{\th_3(\tau)}}
\sqrt{\frac{\th_4(\tau)\th_2(\tau)}{\eta(\tau)^2}}
}~,
\label{Z twisted} 
\end{eqnarray}
where we set $[a]\in \bz_2$, $a\equiv [a]~ (\mod\,2)$, 
and $N^{[S,T]}_{\bl,\tbl}$ are suitably chosen coefficients, which will 
be specified below.  
In the second line we emphasized the $\cN=4$ structure in the
$K3$-sector.
We have adopted an apparent asymmetric form 
as in \cite{HelW}, which seems natural if we recall
$\sigma_{\msc{mirror},R} \sim  \sigma^{\cN=4}_{1,R}$, 
$\sigma_{\msc{mirror},L} \sim \sigma^{\cN=4}_{3,L}$ 
when acting on the $\cN=4$ SCA.


Let us further elaborate contributions from each sector.

~

\noindent
{\bf [1] $S^1$-sector (bosonic) : }
The bosonic part of the $S^1$-direction is represented by the functions 
$Z_{R,(a,b)}(\tau,\bar{\tau})$, $(a,b \in \bz)$. 
Sectors with $a \in 2\bz+1$ or $b \in 2\bz+1$ correspond to twisted
sectors, while contributions from $a,b \in 2\bz$ are included in 
the partition function of the untwisted sector $Z^{\msc{u}}(\tau,\bar{\tau})$.

~


\noindent
{\bf [2] $K3$-sector : } 
As discussed above, the chiral blocks are written in the form of
\begin{eqnarray}
 && \sum_{\bl,\tbl}\, N^{[[a],[b]]}_{\bl,\tbl}\,
\chi^{\bk}_{\bl,[[a],[b]]}(\tau) \overline{\chi^{\bk}_{\tbl,[[a],[b]]}(\tau)}
\equiv \sum_{\bl,\tbl}\, N^{[[a],[b]]}_{\bl,\tbl}\,
f^{\bk}_{\bl,[[a],[b]]}(\tau) \overline{f^{\bk}_{\tbl,[[a],[b]]}(\tau)}
\left|\frac{2}{\th_{[[a],[b]]}(\tau)}\right|^2~, \nn
 && \th_{[0,1]} \equiv \th_2~, ~~ \th_{[1,0]} \equiv \th_4~,~~
 \th_{[1,1]} \equiv \th_3~. 
\label{K3 contribution}
\end{eqnarray}
The relation between the spin structure and the
$\sigma_{\msc{mirror}}$-twisting is slightly non-trivial, and is
summarized in Table 1. 
As was already illustrated, the chiral block for boundary 
condition $[0,1]$ ($a\in 2\bz$, $b\in 2\bz+1$) includes only 
the $\NS$ and $\tNS$-sectors, contributing the same character
function $\chi^{\bk}_{\bl,[0,1]}(\tau)$.
There is no contribution from the $\R$-sector for this boundary condition.


~

\begin{center}
 \begin{tabular}{|c|c|c|c|}\hline
    & [0,1] & [1,0]& [1,1] \\\hline\hline
 NS    & $\chi^{\bk}_{\bl, [0,1]}(\tau) $  & $\chi^{\bk}_{\bl, [1,0]}(\tau)$   
&  0       \\ \hline
 $\tNS$& $\chi^{\bk}_{\bl, [0,1]}(\tau)$  & 0    
&  $ \chi^{\bk}_{\bl, [1,1]}(\tau)$     \\ \hline
 R    &  0  & $\chi^{\bk}_{\bl, [1,0]}(\tau)$    
&  $ \chi^{\bk}_{\bl, [1,1]}(\tau)$     \\ \hline
 $\tR$    &  0  & 0    &  0     \\ \hline
  \end{tabular} \\
\vspace{10pt}
{\bf Table 1.} Relation between the $\sigma_{\msc{mirror}}$-twists and
 the spin structures. \\ 
 Note that $f^{\bk}_{\bl, [\al,\beta]}(\tau)$ are subject to the same relations.   
\end{center}
 
~


The blocks \eqn{K3 contribution} are clearly modular covariant with respect
to the indices $a$, $b$, but the modular transformations generate non-trivial 
mixing of the quantum numbers $\bl$ and $\tbl$.
We thus have to choose the coefficients $N^{[[a],[b]]}_{\bl,\tbl}$ carefully. 
This is accomplished by requiring (in addition to the modular invariance) that 
the orbifold projection 
$\frac{1+\sigma_{\msc{mirror}}}{2}$
acts correctly on the total Hilbert space.
To this aim it is convenient to classify 
the $K3$ Gepner models into the following two cases\footnote
{There is an analogous discussion on modular invariance in \cite{ES-G2orb}.}.

\noindent
{\bf (i) At least one of $k_i$'s is odd}

It is easiest to look at the $[0,1]$-sector
($\sigma_{\msc{mirror}}$-insertion). 
The problem translates into finding out terms that survive the 
$\sigma_{\msc{mirror}}$-insertion in the trace out of the spectral flow orbits
\begin{eqnarray}
 \{(\ell_1,0),\ldots, (\ell_r,0)\}_L\,\otimes \, 
\{(\tilde\ell_1,2n),\ldots, (\tilde\ell_r,2n)\}_R ~~~ (n\in \bz_N)~.
\label{l orbit}
\end{eqnarray}
Under the assumption on $k_i$, we see that only the
terms of the form 
\begin{eqnarray}
 && \prod_i \chi^{k_i}_{\ell_i,[0,1]}(\tau) 
\overline{\chi^{k_i}_{\ell_i,[0,1]}(\tau)}
\label{term 1}
\end{eqnarray}
do survive.
We thus obtain
\begin{eqnarray}
 N^{[[a],[b]]}_{\bl,\tbl}= \prod_{i=1}^r\delta_{\ell_i,
\tilde{\ell}_i}~.
\label{N a b 1}
\end{eqnarray}
This renders \eqn{K3 contribution} trivially modular covariant.

 ~

\noindent
{\bf (ii) All $k_i$'s are even : }

The situation is more involved in this case.
We now have $N \in 2\bz$. 
We define,
\begin{eqnarray}
 && S_1 = \Big\{i\in \{1,\ldots, r\} ~;~ \frac{N}{k_i+2}\in 2\bz+1\Big\}~, \nn
 && S_2 = \Big\{i\in \{1,\ldots, r\} ~;~ \frac{N}{k_i+2}\in 2\bz\Big\}~.
\end{eqnarray}
One finds that, in addition to \eqn{term 1}, terms like
\begin{eqnarray}
 && \prod_{i} \chi^{k_i}_{\ell_i,[0,1]}(\tau) 
\cdot 
\prod_{i\in S_2} 
\overline{\chi^{k_i}_{\ell_i,[0,1]}(\tau)}
\prod_{i\in S_1}
\overline{\chi^{k_i-\ell_i}_{\ell_i,[0,1]}(\tau)}
\label{term 2}
\end{eqnarray}
also contribute (they appear as the $n=N/2$ component 
in the orbit \eqn{l orbit}). 
We thus obtain
\begin{eqnarray}
 && N^{[0,1]}_{\bl,\tbl} = \prod_{i\in S_2}
\delta_{\ell_i,\tell_i} 
\prod_{i\in S_1}\left(\delta_{\ell_i,\tell_i}
+\delta_{\ell_i,k_i-\tell_i}\right) ~,
\label{N a b 2}
\end{eqnarray}
and by taking the modular transformations,
also find 
\begin{eqnarray}
&& N^{[1,0]}_{\bl,\tbl}= N^{[1,1]}_{\bl,\tbl}=
\left(1+(-1)^{\sum_{i\in S_1}\ell_i}\right)
\prod_{i=1}^r\delta_{\ell_i,\tell_i}~.
\label{N a b 3}
\end{eqnarray}
To check the modular covariance we further have to classify
\begin{description}
 \item[(ii)-(a) : ]~  $N\in 4\bz$

In this case\footnote{
Both $N\in 4\bz$ and $N\in 4\bz+2$ are possible for the $K3$ Gepner model,
in contrast to the $CY_3$ case where we only have the first possibility 
$N\in 4\bz$ when all $k_i$ are even\cite{ES-G2orb}.
}  
   we can prove that  
   (1) $S_1\neq \emptyset$, ~
   (2) $\sharp S_1 \in 2\bz$, ~ 
   (3) $k_i\in 4\bz+2$ for ${}^{\forall}i \in S_1$.
 \item[(ii)-(b) : ]~ $N \in 4\bz+2$

This time we have
(1) $S_1\neq \emptyset$, ~ 
(2) $k_i\in 4\bz$ for ${}^{\forall}i \in S_1$.
 
\end{description}
Making use of these properties and the modular transformation 
formulas \eqn{modular twisted minimal},
one can confirm that \eqn{N a b 2}, \eqn{N a b 3}
assure the modular covariance of \eqn{K3 contribution}.

~


\noindent
{\bf [3] The free fermion part : }
The free fermion part of the flat space-time (tranverse part of $\br^{4,1}$)
and the $S^1$-direction consists of four fermions. 
As shown in \eqn{Z twisted}, the partition sums of the free fermion part 
$Z^f_{(a,b)}(\tau,\bar{\tau})$ are given as 
\begin{equation}
Z^f_{(a,b)}(\tau,\bar{\tau})= \left\{
\begin{array}{ll}
\left|
\left(\frac{\th_3}{\eta}\right)^2- 
(-1)^{\frac{a}{2}}\left(\frac{\th_4}{\eta}\right)^2
\right|^2~,
& (a\in 2\bz~,~~ b\in 2\bz+1)~, \\
\left|
\left(\frac{\th_3}{\eta}\right)^2- 
(-1)^{\frac{b}{2}}\left(\frac{\th_2}{\eta}\right)^2
\right|^2~, 
& (a\in 2\bz+1~,~~ b\in 2\bz)~, \\
\left|
\left(\frac{\th_4}{\eta}\right)^2+ 
i(-1)^{\frac{a+b}{2}}\left(\frac{\th_2}{\eta}\right)^2
\right|^2~, 
& (a\in 2\bz+1~,~~ b\in 2\bz+1)~.
\end{array}\right.
\label{Z f a b}
\end{equation}
One may identify, for instance in the $a\in 2\bz$, $b\in 2\bz+1$ sector, 
$\left(\frac{\th_3}{\eta}\right)^2$ is the NS contribution, while 
$(-1)^{\frac{a}{2}}\left(\frac{\th_4}{\eta}\right)^2$ lies in 
the $\tNS$ sector. 
The absence of R sector is due to the structure of the chiral blocks in the $K3$-sector 
(see Table 1).
The terms in the other sectors may be identified similarly. 
It should be remarked that the {\em winding dependent} phase factors
$(-1)^{\frac{a}{2}}$, $(-1)^{\frac{b}{2}}$
and $i(-1)^{\frac{a+b}{2}}$ are necessary 
for the expected modular covariance.  
Indeed, {\em with these phase factors} $Z^f_{(a,b)}(\tau,\bar{\tau})$ behave 
covariantly under the modular transformations,
\begin{eqnarray}
 && Z^f_{(a,b)}(-1/\tau,-1/\bar{\tau}) 
= Z^f_{(b,-a)}(\tau,\bar{\tau})~, ~~~
 Z^f_{(a,b)}(\tau+1,\bar{\tau}+1) 
= Z^f_{(a,a+b)}(\tau,\bar{\tau})~. 
\end{eqnarray}
This can be checked by rewriting the functions $Z^f_{(a,b)}(\tau,\bar{\tau})$
in a unified manner,
\begin{eqnarray}
 && Z^f_{(a,b)}(\tau,\bar{\tau}) =  
\left| \frac{\eta(\tau)^2}{\th\cdot \th_{[a,b]}(\tau)} \cdot
G_{(a,b)}(\tau)\right|^2~, \nn
&& G_{(a,b)}(\tau) \equiv 2 q^{\frac{a^2}{8}}
e^{\frac{i\pi}{4}ab} 
\left(\frac{\th_1(\tau,\frac{a\tau+b}{4})}{\eta(\tau)}\right)^4~,
\label{G a b} \\
&&
\th \cdot \th_{[a,b]}(\tau)
\equiv
\left\{
\begin{array}{ll}
 \th_3(\tau)\th_4(\tau),& ~~ (a \in 2\bz,~ b\in 2\bz+1), \\
 \th_2(\tau)\th_3(\tau),& ~~ (a \in 2\bz+1,~ b\in 2\bz) ,\\
 \th_4(\tau)\th_2(\tau),& ~~ (a \in 2\bz+1,~ b\in 2\bz+1), 
\end{array}
\right.
\nonumber
\end{eqnarray}
from which the modular covariance immediately follows.

~

Assembling these results the total partition function 
$Z(\tau,\bar{\tau})=Z^{\msc{u}}(\tau,\bar{\tau})+Z^{\msc{t}}(\tau,\bar{\tau})$
is indeed verified to be modular invariant 
with the above coefficients $N^{[S,T]}_{\bl,\tbl}$.

~


We conclude this section with several comments.

\noindent
{\bf 1. } By our construction $\sigma_{\msc{mirror},R}$ yields 
the automorphism $\sigma^{\cN=2}_{(i)}$ in each $\cN=2$ minimal sector $M_{k_i}$, 
whereas $\sigma_{\msc{mirror},L}$
does {\em not} induce any automorphism in $M_{k_i}$. 
We also point out that the transformation interpolating 
between $\sigma^{\cN=4}_{3}$ and 
$\sigma^{\cN=4}_{1}$ is generically just an {\em outer} automorphism
of $\cN=4$ SCA. 
Thus there is no self-evident principle {\em a priori} that relates the action of 
$\sigma^{\cN=4}_{3}$ with that of $\sigma^{\cN=4}_{1}$ on the $\cN=4$ 
{\em primary states.} 

~

\noindent
{\bf 2. } As addressed above, the chiral blocks of the left-mover have 
formally the same forms as the right-mover.
This means that if we only look at the closed string spectrum the model 
is indistinguishable from the symmetric orbifold with $\bz_2$-twisting 
$\sigma_L=\sigma_R$, as the closed string partition functions are equal.
Nevertheless, it should be emphasised that the asymmetric 
orbifold (with twist $\sigma_{\msc{mirror},L} \neq \sigma_{\msc{mirror},R}$)
is different from the corresponding symmetric orbifold; the distinction being
crucial for the physics of D-branes in this string vacuum.
As observed in \cite{KawaiS1}, an asymmetric orbifold generally yields different spectrum
of geometric D-branes realized by linear gluing conditions from that of a symmetric type orbifold.
We also point out that the mirror-involution 
$\sigma_{\msc{mirror}}$ given above still includes 
a phase ambiguity due to an ambiguity of 
$\sigma^{\cN=2,(i)}$ in each $M_{k_i}$ sector. 
This phase ambiguity does not affect 
the torus partition function we have obtained, 
but it would become important when we examine the 
D-brane spectrum. 
We hope to report on detailed aspects 
on D-branes in mirrorfolds elsewhere.

~

\noindent
{\bf 3. } These mirrorfold string vacua break the space-time SUSY completely, 
yielding a non-vanishing cosmological constant at the one-loop
level. 
In fact, as is seen in \eqn{Z twisted}, the partition function in the twisted sector
does not vanish at all, in contrast to the untwisted sector which is kept 
supersymmetric. 
In other words, the space-time SUSY is broken by the winding string modes, 
which is similar to the Scherk-Schwarz type $S^1$-compactification \cite{SS}. 
It is easy to see that the most tachyonic winding mode appears in the
sector of $a=1$, which has a mass squared proportional to\footnote
   {The temporal winding $b$ may be dualized to the KK momentum 
     by Poisson resummation in the standard way, and it does not play
     any important role in this context. 
     We also find that the sectors of $a\in 4\bz+2$, $b\in 2\bz+1$ include other
     candidates of winding tachyons because of the wrong GSO
     projection due to the phase factors appearing in \eqn{Z f a b}. 
     However, they are found to be always less tachyonic
     than that of the $a=1$ sector and do not alter the discussion here.} 
\begin{eqnarray}
 h-\frac{1}{2} & = & -\frac{1}{2} 
+ \mbox{min}_{\ell_1,\ldots, \ell_r} \, \left\lb
\sum_{i=1}^r 
h^{t}_{\ell_i} \right\rb+ \frac{R^2}{4} \nn
&\geq &
-\frac{1}{4} + \frac{R^2}{4}~,
\label{winding tachyon mass}
\end{eqnarray}
where $h^t_{\ell_i} \equiv 
\frac{k_i-2 + (k_i-2\ell_i)^2}{16(k_i+2)}+\frac{1}{16}$
are the conformal weights of the twisted characters 
$\chi^{k_i}_{\ell_i,[1,0]}(\tau)$ \eqn{h t}.
(The minimum value of $h^t_{\ell_i}$ is achieved when 
$\ell_i= \left\lb \frac{k_i}{2}\right\rb$, and 
the inequality \eqn{winding tachyon mass}
is saturated iff all the levels $k_i$ are even.)
Thus there is no tachyonic instability as long as $R>1$ (self-dual radius).
Of course, since $R$ is a closed string modulus, 
the base circle may shrink to $R<1$ and in that case we encounter an instability 
due to the winding tachyon condensation. 

~


\noindent
{\bf 4. } 
The modular invariant constructed above is, contrary to what would naively be anticipated, 
{\em not\/} of an order 2 orbifold but rather of an order 4 orbifold. 
This arises from the fact that the free fermion part 
$Z^f_{(a,b)}(\tau,\bar{\tau})$ \eqn{Z f a b} is $\bz_4$-periodic with respect to 
the windings $a$, $b$, rather than $\bz_2$. 
This is related to the {\em chiral\/} spin structures that are characteristic to the type II vacua. 
If instead considering the type 0 vacua, the free fermion part would take a simpler form
\begin{eqnarray}
 && Z^{f, \msc{type 0}}(\tau,\bar{\tau}) \propto 
\frac{1}{2}\left\lb \left|\frac{\th_3}{\eta}\right|^4+ 
 \left|\frac{\th_4}{\eta}\right|^4+ \left|\frac{\th_2}{\eta}\right|^4
\right\rb \left(\equiv Z_{SO(4)_1}(\tau,\bar{\tau})\right)~,
\end{eqnarray}
which has no dependence on the windings $a$, $b$.
Consequently, the type 0 vacua of the mirror-folds are realized as 
order 2 orbifolds as expected from the intuitive picture.

~


\noindent
{\bf 5. } In the above construction we have chosen the first candidate
\eqn{sigma primary L 1} of the left action. 
It is of course natural to ask what would happen if we instead use the 
second candidate of operation \eqn{sigma primary L 2}, which might give rise to an
asymmetric modular invariant, if consistent at all.
We were not able to construct a modular invariant using 
\eqn{sigma primary L 2} for compact $K3$ fibrations, and while we have not
exhausted all possibilities, such a construction seems quite unlikely.  
For instance, in the $a\in 2\bz$, $b\in 2\bz+1$ sector, 
the partition function typically includes terms like
\begin{eqnarray}
 \sum_{\bl,\tbl} N^{[0,1]}_{\bl,\tbl} \left(
\delta_{\bl,{\bf 0}}\, \ch{\cN=4,(\sNS)}{0}(\ell=0;\tau)+
\sum_{n\geq 1} a_{n,\bl} \, 
\ch{\cN=4,(\sNS)}{}(p_{n,\bl};\tau)\right) \cdot 
\overline{
\sum_{n\geq 0} a_{n,\tbl} \, \chi_{[0,1]}(p_{n,\tbl};\tau)
}~.
\end{eqnarray}
The first term in the left moving part is the $\cN=4$ massless character of
spin 0 (graviton character), while the second term consists only of 
the massive characters. 
Due to an involved behavior of the massless character under the 
modular S-transformation, the modular invariance of the total partition function
is likely to be spoiled (see Appendix A).

~


\section{Mirrorfolds with Non-compact K3 Fibrations}

In this section we discuss an extension of the mirrorfold model to include 
{\em non-compact} Gepner-like models in the $K3$-fiber. 
In contrast to the compact fiber case, both \eqn{sigma primary L 1}
and \eqn{sigma primary L 2} are found to be compatible with modular 
invariance.

~

\subsection{Non-SUSY Vacua : Symmetric Modular Invariants}

Let us first consider orbifolding by the twist \eqn{sigma primary L 1} 
as in the previous section. 
We now assume non-compact Gepner-like models 
for the $K3$-fiber, defined by 
\begin{eqnarray}
&& \cM_{\msc{fiber}} \equiv 
\left\lb M_{k_1}\otimes \cdots \otimes M_{k_r} \otimes 
  L_{\bar{N},\bar{K}}\right\rb/ \bz_N~, ~~~ \label{nc Gepner} \\
&& N \equiv \mbox{L.C.M.} \, \{k_i+2, \, \bar{N}\} ~, \nn
&& \sum_{i=1}^r \frac{k_i}{k_i+2} + \left(1+ \frac{2\bar{K}}{\bar{N}}\right)=2~,
\label{nc crit cond}
\end{eqnarray}
where $L_{\bar{N},\bar{K}}$ denotes the $SL(2;\br)/U(1)$ Kazama-Suzuki 
supercoset model at level $k\equiv \bar{N}/\bar{K}$ (for simplicity we assume 
$\bar{N}$ and $\bar{K}$ to be relatively prime hereafter).
Note, in particular, that this includes the $A_{N-1}$-type ALE spaces \cite{OV}
as a simplest case of the fiber SCFT,
$$
\cM_{\msc{fiber}}= \lb M_{N-2} \otimes L_{N,1} \rb/\bz_N~.
$$
With these fibre models we may construct mirrorfolds in the same way as in the previous section,
$$
 \frac{\cM_{\msc{fiber}} \times S^1_{2R}}{\sigma_{\msc{mirror}}\otimes
 \cT_{2\pi R}}~.
$$

Now let us work on the partition function. 
The total partition function generically has the twisted and untwisted parts,
\begin{eqnarray}
 Z(\tau,\bar{\tau}) = Z^{\msc{u}}(\tau,\bar{\tau})+
  Z^{\msc{t}}(\tau,\bar{\tau})~,
\end{eqnarray}
where we again define the twisted sector
$Z^{\msc{t}}(\tau,\bar{\tau})$ as including temporal or spatial  
twist by $\sigma_{\msc{mirror}}\otimes \cT_{2\pi R}$.
The untwisted sector $Z^{\msc{u}}(\tau,\bar{\tau})$ involves no such twist.
We shall discuss each sector separately. 

~

\noindent
{\bf The untwisted sector}

 The partition function $Z^{\msc{u}}(\tau,\bar{\tau})$ 
of the untwisted sector is known to be IR-divergent, 
reflecting the infinite volume of
the non-compact target space. 
The regularized partition function splits into two parts 
\cite{ES-BH} (see also \cite{HPT,IKPT}), 
\begin{eqnarray}
 && Z^{\msc{u}}(\tau,\bar{\tau}) =
  Z^{\msc{u}}_{\msc{con}}(\tau,\bar{\tau})
 + Z^{\msc{u}}_{\msc{dis}}(\tau,\bar{\tau})~,
\end{eqnarray} 
where the first term includes continuous representations of
$L_{\bar{N},\bar{K}}$ and is expanded only with the $\cN=4$ massive characters. 
This is manifestly modular invariant and is proportional to the (regularized) volume 
factor $V \sim \ln \ep$ ($\ep$ is the IR cut-off). 
The continuous part describes the propagating degrees of freedom in the 
non-compact $K3$ space. 
The second term $Z^{\msc{u}}_{\msc{dis}}(\tau,\bar{\tau})$, on the other hand, 
includes discrete representations of $L_{\bar{N},\bar{K}}$. 
There is no volume factor in the second term as it corresponds to the localized 
degrees of freedom around isolated singularities in the background. 
The $\cN=4$ character expansion of the second (discrete) part involves both the 
massless {\em matter\/} characters (i.e. $\ell=1/2$ for the NS sector) and the massive
characters, but {\em no graviton (identity) character}.
This means that gravity decouples in the string vacua. 
A potential problem is that $Z^{\msc{u}}_{\msc{dis}}(\tau,\bar{\tau})$ is 
not modular invariant in general.
A way to circumvent this problem is to focus only on the propagating 
degrees of freedom, by considering the partition function
{\em per unit volume} as discussed in \cite{ES-BH};
\begin{eqnarray}
 \lim_{V\,\rightarrow\, \infty} \frac{Z}{V} = 
\lim_{V\,\rightarrow\, \infty} \frac{Z_{\msc{con.}}}{V}~.
\end{eqnarray}
Note that the second term $Z_{\msc{dis}}$ drops after divided by the
infinite volume factor $V$. 

The partition function for the untwisted sector is obtained as 
 \cite{ncGepner,ES-BH,ES-C}
\begin{eqnarray}
 && \hspace{-25mm}
\frac{Z^{\msc{u}}_{\msc{con.}} (\tau,\bar{\tau})}{V}
= \frac{1}{2}\cdot \frac{1}{4N} \sum_{\al,\tal} \sum_{I,\tI} \, 
\ep(\al)\ep(\tal) N_{I,\tI}\,
\cG^{(\al)}_{I}(\tau)\overline{\cG^{(\tal)}_{\tI}(\tau)}
\, \frac{1}{\tau_2^2\left|\eta(\tau)\right|^8} \,
Z_{2R}(\tau,\bar{\tau}) \left(\frac{\th_{\lb \al \rb}}{\eta}\right)^3
 \overline{\left(\frac{\th_{\lb \tal \rb}}{\eta}\right)^3}~,
\label{Z u nc}
\end{eqnarray}
where the chiral blocks in the NS sector are
\begin{eqnarray}
 && \cG^{(\sNS)}_{I}(\tau,z) \equiv 
\frac{1}{N}
\sum_{a,b\in\bsz_N} q^{\frac{a^2}{2}}y^{a} \prod_{i=1}^r
\ch{(\sNS),k_i}{\ell_i,m_i}(\tau,z+a\tau+b)
\,
\frac{\Th{\bar{m}}{\bar{N}\bar{K}}\left(\tau,\frac{2}{\bar N}(z+a\tau+b)\right)}
{\eta(\tau)}~,
\label{F nc}
\end{eqnarray}
and those for the other spin structures are obtained by the $1/2$ 
spectral flows. 
Here $I \equiv \left\{ (\ell_i,m_i),\, \bar{m}\right\}$ is 
the collective index.
The right-moving chiral blocks are similar, and
$N_{I,\tI}$ are some coefficients (not specified
explicitly here)\footnote
{Note that the quantum numbers $\bar{m}$ in the 
$L_{\bar{N},\bar{K}}$-sector need not be symmetric. 
A typical modular invariant includes 
$$
\bar{m}= \bar{K}n_0+ \bar{N}w_0~, ~~~ \tilde{\bar{m}}= \bar{K}n_0-
\bar{N}w_0~, ~~~ (n_0 \in \bz_{\bar{N}}, w_0 \in \bz_{2\bar{K}})~.
$$
See e.g. \cite{ES-C} for more details.
} that are compatible with modular invariance.
We used the common abbreviation 
$\cG^{(\al)}_I(\tau)\equiv\cG^{(\al)}_I(\tau,0)$.
The $L_{\bar{N},\bar{K}}$-sector yields
additional free oscillator contributions
$
\frac{1}{\left|\eta\right|^2} 
\left(\frac{\th_{[\al]}}{\eta}\right)
\overline{\left(\frac{\th_{[\tal]}}{\eta}\right)}
$,
and the integral over the zero-mode momentum 
of the non-compact boson (`Liouville mode')
generates one more factor $\tau_2^{-1/2}$.
As addressed above, 
the partition function $Z^{\msc{u}}_{\msc{con.}}$
includes only the continuous representations 
in the $L_{\bar{N},\bar{K}}$-sector, and 
its modular invariance is manifest.

~


\noindent
{\bf The twisted sector}

The construction of the twisted sector partition function is similar to the compact case.
We find, 
\begin{eqnarray}
&& \hspace{-2cm}
\frac{Z^{\msc{t}}(\tau,\bar{\tau})}{V} 
= \frac{1}{4}
\sum_{\stackrel{\scriptstyle a\in 2\bsz}{b\in 2\bsz+1}}
Z_{R,(a,b)}(\tau,\bar{\tau}) 
\frac{1}{\tau_2^2\left|\eta\right|^8} 
\sum_{\bl} \, \chi^{\bk}_{\bl,[0,1]}(\tau)  
\overline{\chi^{\bk}_{\bl,[0,1]}(\tau)} \,
\frac{\th_3\th_4}{\eta^2}
\cdot
\overline{
\sqrt{\frac{2\eta}{\th_2}}
\sqrt{\frac{\th_3\th_4}{\eta^2}}
}
\cdot
Z^f_{(a,b)}(\tau,\bar{\tau})~
\nn
&&
\hspace{-5mm}
+ \frac{1}{4}
\sum_{\stackrel{\scriptstyle a\in 2\bsz+1}{b\in 2\bsz}}
Z_{R,(a,b)}(\tau,\bar{\tau}) 
\frac{1}{\tau_2^2\left|\eta\right|^8} 
\sum_{\bl} \, \chi^{\bk}_{\bl,[1,0]}(\tau)  
\overline{\chi^{\bk}_{\bl,[1,0]}(\tau)} \,
\frac{\th_2\th_3}{\eta^2}
\cdot
\overline{
\sqrt{\frac{2\eta}{\th_4}}
\sqrt{\frac{\th_2\th_3}{\eta^2}}
}
\cdot
Z^f_{(a,b)}(\tau,\bar{\tau})~
\nn
&& \hspace{-5mm}
+ \frac{1}{4}
\sum_{\stackrel{\scriptstyle a\in 2\bsz+1}{b\in 2\bsz+1}}
Z_{R,(a,b)}(\tau,\bar{\tau}) 
\frac{1}{\tau_2^2\left|\eta\right|^8} 
\sum_{\bl} \, \chi^{\bk}_{\bl,[1,1]}(\tau)  
\overline{\chi^{\bk}_{\bl,[1,1]}(\tau)} \,
{\frac{\th_4\th_2}{\eta^2}} 
\cdot
\overline{
\sqrt{\frac{2\eta}{\th_3}}
\sqrt{\frac{\th_4\th_2}{\eta^2}}
}
\cdot
Z^f_{(a,b)}(\tau,\bar{\tau})~
~.
\label{Z t nc} 
\end{eqnarray}
where we set $[a]\in \bz_2$, $a\equiv [a]~ (\mod\,2)$ as before.
We used an abbreviated notation 
$\chi^{\bk}_{\bl}(\tau) \equiv \prod_{i=1}^r \chi^{k_i}_{\ell_i}(\tau)$
and 
$Z^f_{(a,b)}(\tau,\bar{\tau})$ is defined in \eqn{Z f a b}.

Note that $\chi^{\bk}_{\bl}(\tau)$ here plays the same role as 
the function $f^{\bk}_{\bl}(\tau)$ in the compact case. 
Namely, it
appears as the trace over the $\cN=4$ primary states.
Again we have an additional contribution of 
$1/\tau_2^{1/2}|\eta|^2$ from the non-compact 
boson along the linear dilaton direction.
Another difference from the compact case is the
modular invariant coefficients
\begin{eqnarray}
 && N_{\bl,\tbl}^{[0,1]}= N_{\bl,\tbl}^{[1,0]}=N_{\bl,\tbl}^{[1,1]} 
= \prod_{i=1}^r \delta_{\ell_i,\tell_i}~.
\label{eqn:NC-SYM-N}
\end{eqnarray}
In the compact case the conformal blocks are related by formulas like
$$
\ch{(\sNS),k}{\ell,m}(\tau,z)
= \ch{(\sNS),k}{k-\ell,m+k+2}(\tau,z)~,
$$
due to the field identification of the minimal models.
In the non-compact case such a relation is absent for the 
conformal blocks of the $L_{\bar{N},\bar{K}}$-sector,
$\dsp \propto q^*\Th{\bar{m}}{\bar{N}\bar{K}}
\frac{\th_{\lb \al \rb}}{\eta^3}$,
giving rise to the relatively simple coefficients (\ref{eqn:NC-SYM-N}).


~

Here we would like to give some comments on the non-compact mirrorfold model.

~

\noindent
{\bf 1. }
As in the compact fiber case, these string vacua are not
supersymmetric, and we can likewise examine the tachyonic instability. 
A slight difference from the compact case is the existence of mass gap 
$\frac{\bar{K}}{4\bar{N}}$.
We find 
\begin{eqnarray}
 h-\frac{1}{2} & = & -\frac{1}{2} 
+ \mbox{min}_{\ell_1,\ldots, \ell_r} \, \left\lb
\sum_{i=1}^r 
h^{t}_{\ell_i} \right\rb
+ \left(\frac{1}{8} + \frac{\bar{K}}{4\bar{N}} \right)
+ \frac{R^2}{4}  ~.
\label{winding tachyon mass nc}
\end{eqnarray}
Using the criticality condition \eqn{nc crit cond},
we again reach the evaluation
\begin{eqnarray}
h-\frac{1}{2}
 &\geq &
-\frac{1}{4} + \frac{R^2}{4}~.
\label{evaluation mass}
\end{eqnarray}
Therefore, we have no tachyonic instability 
as long as $R>1$.
However, there exists a crucial difference from the compact case:
now the graviton modes are decoupled from the physical Hilbert space, 
implying that the radius $R$ becomes non-normalizable. 
Thus we should regard it as a parameter of the theory 
rather than a dynamical modulus. 
There could still exist normalizable 
closed string moduli corresponding to the massless matter
rep. $\cD^{(\sNS)}_{1/2}$ of $\cN=4$ SCA (see \cite{ES-BH,ES-C} for more details).
However, the corresponding marginal deformations do not affect the mass square of 
the winding tachyon \eqn{evaluation mass}, because they must preserve the $\cN=4$
superconformal symmetry. 
We thus conclude that these non-supersymmetric string vacua are 
stable at the level of perturbative string,
as long as $R$ is chosen to be greater than the self-dual radius.

~


\noindent
{\bf 2. }
For the simplest case $\cM_{\msc{fiber}}=\lb M_{N-2}\otimes
L_{N,1}\rb/\bz_N$, which describes the ALE space of $A_{N-1}$-type \cite{OV}, 
we obtain 
%
\begin{eqnarray}
 &&\frac{Z^{\msc{u}}(\tau,\bar{\tau})}{V} = 
\frac{1}{4}\sum_{\al,\tal} \ep(\al)\ep(\bar{\al}) 
Z_{2R}(\tau,\bar{\tau})
Z_{SU(2)_k}(\tau,\bar{\tau}) \cdot
\frac{1}{\tau_2^{2}\left|\eta\right|^8} 
\left(\frac{\th_{\lb \al \rb}}{\eta}\right)^4
\overline{
\left(\frac{\th_{\lb \tal \rb}}{\eta}\right)^4
}
~,
\label{Z u ALE} \\
&& 
\frac{Z^{\msc{t}}(\tau,\bar{\tau})}{V} = \frac{1}{4}
\sum_{\stackrel{\scriptstyle a\in 2\bsz+1}{\msc{or}~b\in 2\bsz+1}}
Z_{R,(a,b)}(\tau,\bar{\tau}) 
\frac{1}{\tau_2^{2}\left|\eta\right|^8} 
\sum_{\ell}
\chi^{k}_{\ell,\,[[a],[b]]}(\tau)
\overline{\chi^{k}_{\ell,\,[[a],[b]]}(\tau)}
\hZ^f_{(a,b)}(\tau,\bar{\tau})~, 
\label{Z t ALE} 
\end{eqnarray}
where the free fermion part is written as 
\begin{equation}
\hZ^f_{(a,b)}(\tau,\bar{\tau})= \left\{
\begin{array}{ll}
\left|
\left(\frac{\th_3}{\eta}\right)^3
\frac{\th_4}{\eta} - 
(-1)^{\frac{a}{2}}\left(\frac{\th_4}{\eta}\right)^3
\frac{\th_3}{\eta}
\right|^2~, & 
(a\in 2\bz~,~~ b\in 2\bz+1)~, \\
\left|
\left(\frac{\th_3}{\eta}\right)^3 
\frac{\th_2}{\eta}- 
(-1)^{\frac{b}{2}}\left(\frac{\th_2}{\eta}\right)^3
\frac{\th_3}{\eta}
\right|^2~, & 
(a\in 2\bz+1~,~~ b\in 2\bz)~, \\
\left|
\left(\frac{\th_4}{\eta}\right)^3\frac{\th_2}{\eta}
+ i(-1)^{\frac{a+b}{2}}\left(\frac{\th_2}{\eta}\right)^3
\frac{\th_4}{\eta}
\right|^2~, & 
(a\in 2\bz+1~,~~ b\in 2\bz+1)~.
\end{array}\right.
\label{hat Z f a b}
\end{equation}
To derive \eqn{Z u ALE} we have used the familiar branching relation for 
the $\cN=2$ minimal characters \eqn{branching 1}, \eqn{branching 2}. 
We also note 
\begin{eqnarray}
 &&\sum_{\ell}
\chi^{k}_{\ell,\,[[a],[b]]}(\tau)
\overline{\chi^{k}_{\ell,\,[[a],[b]]}(\tau)}
= 
\sum_{\ell}
\chi^{SU(2)_k}_{\ell,\,(a,b)}(\tau)
\overline{\chi^{SU(2)_k}_{\ell,\,(a,b)}(\tau)}~,
\label{relation 1}
\end{eqnarray}
where the R.H.S is written in terms of the twisted $SU(2)_k$ characters 
\eqn{twisted SU(2) ch}. \footnote
  {It is important to notice that $\left|\chi^{SU(2)_k}_{\ell,\,(a,b)}(\tau)\right|^2$ is
  $\bz_2$-periodic with respect to $a$, $b$, even though the chiral part
  $\chi^{SU(2)_k}_{\ell,\,(a,b)}(\tau)$ breaks that periodicity due to an extra 
  phase factor. Therefore, \eqn{relation 1} is a consistent 
relation. }
This is consistent with the fact 
that $\sigma_{\msc{mirror}}$ is now 
interpretable as the $(e^{i\pi K^3_0}, e^{i\pi
\widetilde{K}^1_0})$-twisting in the $SU(2)$ supersymmetric 
WZW model of level $N\equiv k+2$, 
where $K^a$ are the (total) $SU(2)$-currents, 
if recalling \cite{OV} 
$$
\lb M_{N-2}\otimes L_{N,1} \rb/\bz_N \cong \br_{\phi} \times
SU(2)_{N}~. 
$$
Obviously, the model defined by \eqn{Z u ALE}, \eqn{Z t ALE} is regarded as a
supersymmetric analogue of the $SU(2)$ T-fold considered in 
\cite{KawaiS1}.
In this case the interpolation between $\sigma^{\cN=4}_1$ and 
$\sigma^{\cN=4}_3$ is exceptionally realized as an inner automorphism. 


~


\subsection{SUSY Vacua : Asymmetric Modular Invariants}

In contrast to the compact fiber case, the second candidate of
the mirror-involution \eqn{sigma primary L 2} turns out to yield 
consistent mirrorfolds, as we shall demonstrate below.
We denote the involution of \eqn{sigma primary L 2} as 
$\hsigma_{\msc{mirror}}$ in order to distinguish it from the first one. 
Since $\hsigma_{\msc{mirror}}$ acts on the
$\cN=4$ SCA as $({\bf 1}, \sigma^{\cN=4}_{1,R})$, the resultant 
partition function will provide an asymmetric modular invariant. 
What differs crucially from the compact models is that we include {\em only\/}
the massive representations. 
The massive characters possess simpler modular
properties that makes an asymmetric modular invariant possible.

The model is described as follows. 
The untwisted sector has the same partition function \eqn{Z u nc}.
The partition function in the 
twisted sector is given as 
\begin{eqnarray}
&& \hspace{-1cm}
Z^{\msc{t}}(\tau,\bar{\tau}) = \frac{1}{4}
\sum_{\stackrel{\scriptstyle a\in 2\bsz+1}{\msc{or}~b\in 2\bsz+1}}
Z_{R,(a,b)}(\tau,\bar{\tau}) 
\frac{1}{\tau_2^{2}\left|\eta\right|^8} 
\sum_{\bl}
\chi^{\bk}_{\bl,\,[[a],[b]]}(\tau)
\overline{\chi^{\bk}_{\bl,\,[[a],[b]]}(\tau)}
\hZ^{f, \msc{SUSY}}_{(a,b)}(\tau,\bar{\tau})~.
\label{Z t SUSY} 
\end{eqnarray}
The free fermion part is now written as 
\begin{eqnarray}
 && \hZ^{f, \msc{SUSY}}_{(a,b)}(\tau,\bar{\tau}) 
= \left\lb \left(\frac{\th_3}{\eta}\right)^4 
-\left(\frac{\th_4}{\eta}\right)^4
-\left(\frac{\th_2}{\eta}\right)^4
\right\rb \cdot 
\overline{
G_{(a,b)} (\tau)}~,
\end{eqnarray}
where $G_{(a,b)}(\tau)$ is defined in \eqn{G a b}.
More explicitly,
\begin{equation}
\begin{array}{l}
\hZ^{f, \msc{SUSY}}_{(a,b)}(\tau,\bar{\tau})\\
\hspace{-10mm}
= \left\{
\begin{array}{ll}
e^{\frac{i\pi}{4}ab}\,
\left\lb \left(\frac{\th_3}{\eta}\right)^4 
-\left(\frac{\th_4}{\eta}\right)^4
-\left(\frac{\th_2}{\eta}\right)^4
\right\rb
\cdot
\overline
{\left\lb
\left(\frac{\th_3}{\eta}\right)^3
\frac{\th_4}{\eta} - 
(-1)^{\frac{a}{2}}\left(\frac{\th_4}{\eta}\right)^3
\frac{\th_3}{\eta}
\right\rb}
~,
&
(a\in 2\bz, b\in 2\bz+1)~, \\
e^{-\frac{i\pi}{4}ab}\,
\left\lb
\left(\frac{\th_3}{\eta}\right)^4 
-\left(\frac{\th_4}{\eta}\right)^4
-\left(\frac{\th_2}{\eta}\right)^4
\right\rb
\cdot \overline{
\left\lb
\left(\frac{\th_3}{\eta}\right)^3 
\frac{\th_2}{\eta}- 
(-1)^{\frac{b}{2}}\left(\frac{\th_2}{\eta}\right)^3
\frac{\th_3}{\eta}
\right\rb}
~, 
&
(a\in 2\bz+1, b\in 2\bz)~, \\
- e^{-\frac{i\pi}{4}ab}\,
\left\lb
\left(\frac{\th_3}{\eta}\right)^4 
-\left(\frac{\th_4}{\eta}\right)^4
-\left(\frac{\th_2}{\eta}\right)^4
\right\rb
\cdot \overline{
\left\lb
\left(\frac{\th_4}{\eta}\right)^3\frac{\th_2}{\eta}
+ 
i(-1)^{\frac{a+b}{2}}\left(\frac{\th_2}{\eta}\right)^3
\frac{\th_4}{\eta}
\right\rb}
~, 
&
(a\in 2\bz+1, b \in 2\bz+1)~. 
\end{array}\right.\end{array}
\label{hat Z SUSY f a b}
\end{equation}
At first glance, \eqn{hat Z SUSY f a b} might appear 
inconsistent with unitarity due to phase factors 
depending on winding numbers $a$, $b$. 
In a unitary theory the torus partition function 
has to be real when $\tau=i\tau_2$
(i.e.  $\mbox{Re}\, \tau=0$). 
We note that, when $\tau=i\tau_2$, 
\begin{eqnarray}
 && \overline{Z_{R,(a,b)}(\tau,\bar{\tau})}
= Z_{R,(a,-b)}(\tau,\bar{\tau}) ~, ~~~
\overline{\widehat{Z}^{f,\msc{SUSY}}_{(a,b)}(\tau,\bar{\tau})}
= \widehat{Z}^{f,\msc{SUSY}}_{(a,-b)}(\tau,\bar{\tau})~,
\end{eqnarray}
so the total partition function is indeed real when $\tau=i\tau_2$,
after summing over $a$, $b$.
Hence there is no inconsistency with unitarity.

In these models the left-movers are expanded by the $\cN=4$ massive
characters {\em with no twisting}, which contribute
as $\left(\frac{\th_{[\al]}}{\eta}\right)^2$ in the partition sum.
The right-moving chiral blocks are twisted by $\sigma^{\cN=4}_1$,
so the models are interpreted as mirrorfolds. 
Recall that the twisted characters contribute to the free fermion part as 
$\frac{\th\cdot\th_{[a,b]}}{\eta^2}$ (see \eqn{G a b}).
The space-time SUSY is achieved by the standard GSO projection acting
only on the left-mover, which preserves 8 supercharges.


The model is free from any tachyonic instability 
in these supersymmetric mirrorfolds, as it should. 
If we only look at the right-mover, 
it might seem possible to have winding 
tachyon modes (belonging to the $a\in 2\bz+1$-sectors), 
similarly to the previous argument in the compact case.
However, this does not happen because such string 
excitations never satisfy the level matching condition 
and the physical Hilbert space does not include them.

~


\subsection{Comments on D-branes: 
Breakdown of the Space-time SUSY}

Finally, we mention some interesting features of
D-branes in these supersymmetric mirrorfolds, although detailed studies
on D-branes will be left to our future work.

A remarkable fact is that {\em all D-branes in these string
vacua are non-BPS.} Recall that the space-time supercharges only come 
from the left-mover, so no boundary state can preserve the space-time SUSY. 
In other words, adding any D-brane breaks the space-time SUSY completely.

A typical boundary state describing a D-brane in these 
vacua has the form 
\begin{eqnarray}
 \ket{B} = \frac{1+\hsigma_{\msc{mirror}}\otimes \cT_{2\pi R}}
{\sqrt{2}} \, \ket{B}_0~,
\end{eqnarray}
where $\ket{B}_0$ is a boundary state in the `parent theory'
$K3\times S^1_{2R}$. As just mentioned, adding this brane 
breaks the space-time SUSY, so we expect to have
open string tachyons which would lead to an IR instability 
of this vacuum.

Let us briefly discuss whether the cylinder amplitude such as 
$\bra{B}e^{-\pi s H^{(c)}} \ket{B}$ ($H^{(c)}$ is the closed 
string Hamiltonian) gives rise to an IR instability. 
After taking account of the contribution from the 
flat space-time and summing over spin structures, 
the term with no insertion of $\hsigma_{\msc{mirror}}\otimes
\cT_{2\pi R}$ provides a vanishing open string amplitude, 
because the GSO projection correctly acts on it. 
However, this is not the case for
the term in which $\hsigma_{\msc{mirror}}\otimes
\cT_{2\pi R}$ is inserted, due to the lack of GSO projection 
in the open string channel. 
It is not difficult to see that the NS sector yields the leading contribution 
to the non-SUSY piece of the open channel amplitude. 
It would look like 
($q=e^{-2\pi t}$, $t\equiv 1/s$)
\begin{eqnarray}
\hspace{-15mm}
 Z^{\msc{cyl}, (\sNS)}(it) &\sim & \int_0^{\infty} dp 
\sum_I \sum_{n\in \frac{1}{2}\bsz_{\geq 0}} \, \rho_I(p)  
c_{I,n} 
q^{\frac{p^2}{2}+h_I+\frac{\bar{K}}{4\bar{N}} + n - \frac{1}{8} }
\frac{2}{\theta_4(it)} 
\times \left\lb \mbox{sectors other than $K3$}\right\rb
~, 
\end{eqnarray}
with some non-trivial  
density function\footnote
  {In the simple case of ALE fiber, 
   the density $\rho_I(p)$ is explicitly calculated in 
   \cite{NST}. See also \cite{N=2Lbrane}.}
$\rho_I(p)$
and coefficients $c_{I,n} \in \bz_{\geq 0}$ determined from
the boundary wave function of $\ket{B}_0$.
The relevant term contributes to the lightest open string 
mode as $h_{\msc{min}}= \frac{\bar{K}}{4\bar{N}}+\frac{1}{8}$.
Here, the contribution $1/8$ is due to the twisted character
$
\chi_{[1,0]}(p;it) = q^{\frac{p^2}{2}}\frac{2}{\th_4(it)}
$,
(with $h=\frac{p^2}{2}+\frac{1}{4}$).
When the brane is 
localized along the base circle, 
we also have winding energy of open strings $R^2$
originating from the $\cT_{2\pi R}$ insertion, whereas
no more contribution when the brane is wrapped around 
the base. 

To summarise,  
\begin{itemize}
 \item {\bf D-branes localized along the base : }
The open string mass squared behaves as
\begin{eqnarray}
 h-\frac{1}{2} \geq \frac{\bar{K}}{4\bar{N}}-\frac{3}{8}+R^2~.
\end{eqnarray} 
Hence the vacuum is IR stable 
as long as $R>R_c \equiv \sqrt{\frac{3}{8}
-\frac{\bar{K}}{4\bar{N}}}$, whereas unstable if $R<R_c$.
 \item {\bf D-branes wrapped around the base : }
The open string mass squared behaves as
\begin{eqnarray}
 h-\frac{1}{2} \geq \frac{\bar{K}}{4\bar{N}}-\frac{3}{8}~.
\end{eqnarray}
The vacuum is always IR unstable.
(Note that $\frac{\bar{K}}{4\bar{N}}\leq \frac{1}{8}$ 
holds because of the criticality 
condition \eqn{nc crit cond}.)
\end{itemize}
%
Again $R$ is not a normalizable modulus, and any normalizable 
moduli inherited from both closed and open string modes 
do not affect the above evaluation of the lightest open string 
mass.

~


\section{Discussions}

In this paper we have studied a class of non-geometric 
backgrounds of superstring theory defined with the 
twisting by the mirror transformation on a $K3$ space 
which we call `mirrorfolds'. 
We have mainly elaborated on how we can construct modular invariant 
models that describe mirrorfolds.
We have also discussed possible instability caused by winding tachyon condensations. 
To achieve modular invariance, it has been crucial to carefully
fix the action of the mirror-involution on the $\cN=4$ primary
states.

It would be a little surprising that we have
several significant distinctions between the compact and the
non-compact models. As we have demonstrated, 
supersymmetric mirrorfolds can exist only in 
the non-compact models in which gravity decouples.
We have also found that the compact mirrorfolds are always unstable 
due to the tachyonic modes wound around the base circle. 
From the viewpoints of representation theory of $\cN=4$ SCA, 
the difference of these two theories originates from 
the modular properties of the irreducible characters of the
$\cN=4$ SCA. 
The graviton character, which only appears in the compact
models, has complicated modular properties that
makes possibility of modular invariance
so restricted compared with the non-compact models. 

A possible future direction related to the present work
would be to study D-branes in these vacua. 
As we have already mentioned (see the comment 2 at the end
of section 2), the phase ambiguity 
of $\sigma_{\msc{mirror}}$ has not been completely removed. 
This would be important when working with the D-brane spectrum,
although it was immaterial for the construction 
of modular invariant partition functions. 
In particular, how the Cardy conditions restrict this phase ambiguity 
is an interesting issue to study.

It is also interesting to compare the analysis given in this 
paper with models in which the $K3$-fibers
are realized as orbifolds $T^4/\Gamma$, where $\Gamma$
is some discrete subgroup of $SU(2) \subset SO(4)$ 
acting on $T^4$. 
As is familiar \cite{EOTY}, some of Gepner points are also 
interpretable as orbifolds of $T^4$, 
and it will be anticipated that the T-fold construction 
works for those orbifold models. 
It would be non-trivial, however, whether such a T-folding is equivalent
with the `mirrorfolding' argued in this paper, or how these two should
be related. 


~

~

\section*{Acknowledgments}

Y.S. was partly supported by Ministry of Education, Culture, Sports, Science and Technology of Japan.
S.K. acknowledges support from JSPS (Research Fellowship for Young Scientists) and the Academy of Finland (Finnish-Japanese Core Programme, grant 112420).



~

~

\section*{Appendix A: ~ Some Conventions and Notations}

\setcounter{equation}{0}
\def\theequation{A.\arabic{equation}}

In this Appendix we collect formulae frequently used in the paper.
We use modular parameters $q \equiv e^{2\pi i \tau}$, $y\equiv e^{2\pi i z}$ and theta functions defined by
 \begin{equation}
 \begin{array}{l}
 \dsp \th_1(\tau,z)=i\sum_{n=-\infty}^{\infty}(-1)^n q^{(n-1/2)^2/2} y^{n-1/2}
  \equiv 2 \sin(\pi z)q^{1/8}\prod_{m=1}^{\infty}
    (1-q^m)(1-yq^m)(1-y^{-1}q^m), \\
 \dsp \th_2(\tau,z)=\sum_{n=-\infty}^{\infty} q^{(n-1/2)^2/2} y^{n-1/2}
  \equiv 2 \cos(\pi z)q^{1/8}\prod_{m=1}^{\infty}
    (1-q^m)(1+yq^m)(1+y^{-1}q^m), \\
 \dsp \th_3(\tau,z)=\sum_{n=-\infty}^{\infty} q^{n^2/2} y^{n}
  \equiv \prod_{m=1}^{\infty}
    (1-q^m)(1+yq^{m-1/2})(1+y^{-1}q^{m-1/2}), \\
 \dsp \th_4(\tau,z)=\sum_{n=-\infty}^{\infty}(-1)^n q^{n^2/2} y^{n}
  \equiv \prod_{m=1}^{\infty}
    (1-q^m)(1-yq^{m-1/2})(1-y^{-1}q^{m-1/2}) .
 \end{array}
 \end{equation}
 We also use
 \begin{eqnarray}
 \Th{m}{k}(\tau,z)&=&\sum_{n=-\infty}^{\infty}
 q^{k(n+\frac{m}{2k})^2}y^{k(n+\frac{m}{2k})} ,
 \end{eqnarray}
 and the Dedekind function
 \begin{equation}
 \eta(\tau)=q^{1/24}\prod_{n=1}^{\infty}(1-q^n).
 \end{equation}
 We abbreviate as $\th_i \equiv \th_i(\tau, 0)$
 ($\th_1\equiv 0$), $\Th{m}{k}(\tau) \equiv \Th{m}{k}(\tau,0)$
when no confusion arises.
The character of $SU(2)_k$ with spin 
$\ell/2$ ($0\leq \ell \leq k$) is
\begin{equation}
\chi^{SU(2)_k}_{\ell}(\tau, z) =
\frac{\Th{\ell+1}{k+2}(\tau,z)-\Th{-\ell-1}{k+2}(\tau,z)}
               {\Th{1}{2}(\tau,z)-\Th{-1}{2}(\tau,z)}.
\end{equation}

The branching relation corresponding to the coset construction of the $\cN=2$ minimal models
$\frac{SU(2)_k \times U(1)_2}{U(1)_{k+2}}$ is given by
\begin{eqnarray}
&& \chi_{\ell}^{SU(2)_k}(\tau,w)\Th{s}{2}(\tau,w-z)
=\sum_{m\in \bsz_{2(k+2)}} \chi_m^{\ell,s}(\tau,z)\Th{m}{k+2}(\tau,w-\frac{2z}{k+2}).
\label{branching 1}
\end{eqnarray}
Here,
\begin{eqnarray}
\chi_m^{\ell,s}(\tau,z)=\sum_{r\in \bsz_k}c^{(k)}_{\ell,m-s+4r}(\tau)
\Th{2m+(k+2)(-s+4r)}{2k(k+2)}(\tau,\frac{z}{k+2})~,
\end{eqnarray}
$s\in \bz_4$, 
and $c^{(k)}_{\ell,m}(\tau)$ are the level $k$ string functions defined by 
\begin{eqnarray}
 && \chi^{SU(2)_k}_{\ell}(\tau,z) = \sum_{m\in \bsz_{2k}} 
c^{(k)}_{\ell,m}(\tau) \Th{m}{k}(\tau,z)~.
\end{eqnarray}
The $\cN=2$ minimal model characters are related to $\chi_m^{\ell,s}$ as
\begin{eqnarray}
&&  \ch{(\sNS),k}{\ell,m}(\tau,z) 
\equiv \tr_{{\cal H}^{\sNS}_{\ell,m}} q^{L_0-\hat{c}/8}y^{J_0}
 = \chi^{\ell,0}_m(\tau,z) + \chi^{\ell,2}_m(\tau,z) ~,\nn
&&
\ch{(\stNS),k}{\ell,m}(\tau,z)
  \equiv \tr_{{\cal H}^{\sNS}_{\ell,m}} (-1)^F q^{L_0-\hat{c}/8}y^{J_0}
 = \chi^{\ell,0}_m(\tau,z) - \chi^{\ell,2}_m(\tau,z) ~, \nn
&&
\ch{(\sR),k}{\ell,m} (\tau,z)
\equiv \tr_{{\cal H}^{\sR}_{\ell,m}} q^{L_0-\hat{c}/8}y^{J_0}
 = \chi^{\ell,1}_m(\tau,z) + \chi^{\ell,-1}_m(\tau,z) ~, \nn
&&
\ch{(\stR),k}{\ell,m} (\tau,z)
\equiv 
\tr_{{\cal H}^{\sR}_{\ell,m}}(-1)^F q^{L_0-\hat{c}/8}y^{J_0}
 = \chi^{\ell,1}_m(\tau,z) - \chi^{\ell,-1}_m(\tau,z) ~.
\label{branching 2}
\end{eqnarray}

~


The level 1 (small) $\cN=4$ characters are given as \cite{ET}
\begin{description}
 \item[massive characters : ]
\begin{eqnarray}
  \ch{\cN=4,(\sNS)}{}(h;\tau,z) &=& q^{h-\frac{1}{8}}
\frac{\th_3(\tau,z)^2}{\eta(\tau)^3}~, ~~~ (\mbox{for} ~\cC^{(\sNS)}_h)~.
\label{N=4 massive}
\end{eqnarray} 
 \item[massless characters : ]
\begin{eqnarray}
\ch{\cN=4,(\sNS)}{0}(\ell=\frac 12;\tau,z) &=& 
q^{-1/8}\,
\sum_{n\in \bsz}\, \frac{1}{1+yq^{n-1/2}}\, q^{\frac{n^2}{2}}y^n
\frac{\th_3(\tau,z)}{\eta(\tau)^3}~, ~~~ (\mbox{for} ~\cD^{(\sNS)}_{1/2})~,
\label{N=4 massless matter}\\
\ch{\cN=4,(\sNS)}{0}(\ell=0;\tau,z) &=& 
q^{-1/8}\,
\sum_{n\in \bsz}\, 
\frac{yq^{n-1/2}-1}{1+yq^{n-1/2}}\, q^{\frac{n^2}{2}}y^n
\frac{\th_3(\tau,z)}{\eta(\tau)^3} \nn
&=& 
q^{-1/8}\,
\sum_{n\in \bsz}\, \frac{(1-q)q^{\frac{n^2}{2}+n-\frac{1}{2}}y^{n+1}}
{(1+yq^{n+1/2})(1+yq^{n-1/2})} 
\frac{\th_3(\tau,z)}{\eta(\tau)^3}~, ~~~ (\mbox{for}
~\cD^{(\sNS)}_{0})~. \nn
&&
\label{N=4 grav}
\end{eqnarray}
\end{description}
The following identity is often useful:
\begin{eqnarray}
 && \hspace{-5mm}
\ch{\cN=4,(\sNS)}{}(h;\tau,z)= 
q^{h}\left(
\ch{\cN=4,(\sNS)}{0}(\ell=0;\tau,z) +
2\ch{\cN=4,(\sNS)}{0}(\ell=\frac{1}{2};\tau,z)\right) ~.
\label{N=4 ch id}
\end{eqnarray}
An important property of the $\cN=4$ characters is that they decompose into
spectrally flowed  $\cN=2$ irreducible characters \cite{ET},
\begin{eqnarray}
&& \ch{\cN=4, (\sNS)}{} (h;\tau,z) = \sum_{n\in \bz} q^{n^2}y^{2n} 
\ch{\cN=2, (\sNS)}{} (h, Q=0;\tau,z+n\tau)~,\nn
&& \ch{\cN=4, (\sNS)}{0} (\ell=\frac 12;\tau,z) = \sum_{n\in \bz} q^{n^2}y^{2n} 
\ch{\cN=2, (\sNS)}{M} (Q=1;\tau,z+n\tau)~,\nn
&& \ch{\cN=4, (\sNS)}{0} (\ell=0;\tau,z) = \sum_{n\in \bz} q^{n^2}y^{2n} 
\ch{\cN=2, (\sNS)}{G} (\tau,z+n\tau)~,
\label{decomp N=4 ch} 
\end{eqnarray}
where the three types of $\cN=2$ irreducible characters at $\hat{c}=2$
are given as
\begin{description}
 \item[massive characters : ]
\begin{eqnarray}
&& \ch{\cN=2,(\sNS)}{}(h,Q;\tau,z) = q^{h-\frac{1}{8}}y^Q
\frac{\th_3(\tau,z)}{\eta(\tau)^3}~, 
\label{N=2 massive}
\end{eqnarray} 
\item[massless matter characters : ]
\begin{eqnarray}
&& \ch{\cN=2, (\sNS)}{M}(Q;\tau,z) = 
q^{\frac{|Q|}{2}-\frac{1}{8}}y^Q\,
\frac{1}{1+y^{\msc{sgn}(Q)}q^{1/2}}
\frac{\th_3(\tau,z)}{\eta(\tau)^3}~,
\label{N=2 massless matter}
\end{eqnarray}
 \item[graviton character : ]
\begin{eqnarray}
&& \ch{\cN=2,(\sNS)}{G}(\tau,z) = 
q^{-1/8}
\frac{(1-q)q^{-1/2}y}{(1+yq^{1/2})(1+yq^{-1/2})} 
\frac{\th_3(\tau,z)}{\eta(\tau)^3}~.
\label{N=2 grav}
\end{eqnarray}
\end{description}

The R-sector characters are obtained by the 1/2-spectral flow. Namely, 
\begin{eqnarray}
 && \ch{\cN=4,(\sR)}{}(h;\tau,z) = q^{\frac{1}{4}}y \, \ch{\cN=4,(\sNS)}{}
(h-\frac{1}{4};\tau,z+\frac{\tau}{2})~, ~~~ (\mbox{for}~ \cC^{(\sR)}_h)~, \nn
 && \ch{\cN=4,(\sR)}{0}(\ell;\tau,z) = q^{\frac{1}{4}}y \, \ch{\cN=4,(\sNS)}{0}
(\frac{1}{2}-\ell;\tau,z+\frac{\tau}{2})~, ~~~ (\mbox{for}~ \cD^{(\sR)}_{\ell})~. 
\label{N=4 R ch}
\end{eqnarray}

~


For the convenience of readers we also reproduce the modular transformation 
formulas of the $\cN=4$ characters at level 1 \cite{ET}. 
We only give the NS sector results as the others are readily obtained by spectral flows.
\begin{description}
 \item[(i)~ massive representations]
\begin{eqnarray}
&& \hspace{-1.5cm}
\ch{\cN=4,(\sNS)}{}\left(h=\frac{p^2}{2}+\frac{1}{8};-\frac{1}{\tau},
\frac{z}{\tau}\right) 
=2e^{i\pi\frac{2 z^2}{\tau}}\,
\int_{0}^{\infty}dp'\, \cos(2\pi pp')\,
\ch{\cN=4,(\sNS)}{}\left(h=\frac{{p'}^2}{2}+\frac{1}{8};\tau,z\right)~,
\nn
&&
\label{S N=4 massive}
\end{eqnarray}
\item[(ii)~ massless representations]
\begin{eqnarray}
&& \ch{\cN=4,(\sNS)}{0}\left(\ell=0;-\frac{1}{\tau}, \frac{z}{\tau}\right)
= e^{i\pi \frac{2 z^2}{\tau}}\,\left\{
2\ch{\cN=4,(\sNS)}{0}(\ell=\frac 12;\tau,z)   \right. \nn
&&\hspace{1.5cm}
\left. + 2 \int_{0}^{\infty}dp'\, \sinh(\pi p')\tanh(\pi p')\,
\ch{\cN=4,(\sNS)}{}\left(h=\frac{{p'}^2}{2}+\frac{1}{8};\tau,z\right)
\right\} ~,
\label{S N=4 massless 1}\\
&& \ch{\cN=4,(\sNS)}{0}\left(\ell=\frac 12;-\frac{1}{\tau}, \frac{z}{\tau}\right)
= e^{i\pi \frac{2 z^2}{\tau}}\,\left\{
-\ch{\cN=4,(\sNS)}{0}(\ell=\frac 12;\tau,z)   \right. \nn
&&\hspace{1.5cm}
\left. + 
\int_{0}^{\infty}dp'\, \frac{1}{\cosh(\pi p')}\,
\ch{\cN=4, (\sNS)}{}\left(h=\frac{{p'}^2}{2}+\frac{1}{8};\tau,z\right)
\right\} ~.
\label{S N=4 massless 2}
\end{eqnarray}
\end{description}
Note the appearance of {\em both\/} continuous and discrete 
terms in the massless formulas \eqn{S N=4 massless 1} and \eqn{S N=4
massless 2}. This feature is characteristic to the massless representations.

~


\section*{Appendix B ~ Twisted Characters of $\cN=2$ SCFT}
\setcounter{equation}{0}
\def\theequation{B.\arabic{equation}}

~

The twisted $\cN=2$ characters are defined with respect to
$\bz_2$-autormorphism of the $\cN=2$ SCA,
\begin{equation}
\sigma^{\cN=2} ~:~ T\,\longrightarrow\, T,~~~J\,\longrightarrow\, -J,~~~
G^{\pm}\,\longrightarrow\, G^{\mp} ~. 
\label{sigma twist}
\end{equation}
We denote the twisted characters as 
$\ch{(\al)}{[S,T]}$, where
$\al$ are the spin structures, and $S,T\in \bz_2$ signify the spatial 
and temporal boundary conditions associated with the $\sigma^{\cN=2}$-twist 
($S,T=1$ means twisted, and $S,T=0$ means no twist).
As the $\sigma^{\cN=2}$-twist projects out states with non-vanishing $U(1)$-charges, 
the conformal weights are the only quantum numbers relevant in the twisted sectors.
It is easy to verify the following identities (see e.g.
\cite{ES-G2orb}):
\begin{eqnarray}
&& \ch{(\sNS)}{[0,1]}(\tau)= \ch{(\stNS)}{[0,1]}(\tau)~, ~~~
\ch{(\sNS)}{[1,0]}(\tau) = \ch{(\sR)}{[1,0]}(\tau)~, 
~~~ \ch{(\stNS)}{[1,1]}(\tau)= \ch{(\sR)}{[1,1]}(\tau) ~, 
\label{group 1} \\
&&  \ch{(\sR)}{[0,1]}(\tau)= \ch{(\stR)}{[0,1]}(\tau)~, ~~~
\ch{(\stNS)}{[1,0]}(\tau) = \ch{(\stR)}{[1,0]}(\tau)~, 
~~~ \ch{(\sNS)}{[1,1]}(\tau)= \ch{(\stR)}{[1,1]}(\tau) ~ .
\label{group 2}
\end{eqnarray}
%
We denote the twisted characters in the first line \eqn{group 1}
as $\chi_{[0,1]}(\tau)$, $\chi_{[1,0]}(\tau)$ and 
$\chi_{[1,1]}(\tau)$.
%
To find their explicit forms, it is easiest to first evaluate the characters $\chi_{[0,1]}\equiv
\tr [\sigma^{\cN=2} q^{L_0-\frac{\hat{c}}{8}}]$ and then modular
transform them to the other boundary conditions. 
It is obvious that only neutral $(Q=0)$ representations that are invariant under
$\sigma^{\cN=2}$-action can contribute to these characters. 
For any $\cN=2$ SCFT with $\hat{c}>1$, they are written in simple forms,
\begin{eqnarray}
&&\chi_{[0,1]}(p;\tau) = \frac{2 q^{\frac{p^2}{2}}}{\th_2(\tau)}~, ~~~
(h= \frac{p^2}{2} + \frac{\hat{c}-1}{8})~, \nn
&&\chi_{[1,0]}(p;\tau) = \frac{2 q^{\frac{p^2}{2}}}{\th_4(\tau)}~, ~~~
(h= \frac{p^2}{2} + \frac{\hat{c}}{8})~, \nn
&&\chi_{[1,1]}(p;\tau) = \frac{2 q^{\frac{p^2}{2}}}{\th_3(\tau)}~, ~~~
(h= \frac{p^2}{2} + \frac{\hat{c}}{8})~. 
\label{twisted massive}
\end{eqnarray} 

For the second line \eqn{group 2}, only the representations that are kept 
invariant under $\sigma^{\cN=2}$ can again contribute to $\ch{(\sR)}{[0,1]}$ (or $\ch{(\stR)}{[0,1]}$). 
Most of such representations, however, yield vanishing characters due to fermionic
zero-modes. 
There only exists one exception: the representations generated
by Ramond ground states ($h=\frac{\hat{c}}{8}$) with $Q=0$. In that case, 
oscillator parts cancel out (as in Witten index), and we simply obtain 
\begin{eqnarray}
 \ch{(\sR)}{[0,1]}(h=\frac{\hat{c}}{8}, Q=0;\tau)
\left(= \ch{(\stR)}{[0,1]}(h=\frac{\hat{c}}{8}, Q=0;\tau)\right)
= \pm \frac{\dsp \prod^{\infty}_{n=1}(1+q^n)(1-q^n)}{\dsp
\prod_{n=1}^{\infty}(1+q^n)(1-q^n)}
=\pm 1~.
\label{R 0 1 ch}
\end{eqnarray}
Here we have a sign ambiguity depending on the $\sigma^{\cN=2}$-action
on Ramond ground states. 
The characters of the other boundary conditions  in \eqn{group 2} 
are easily obtained by modular transformations; they are merely equal
$\pm 1$.


The twisted characters of the minimal models $M_{k}$ 
are more involved.
The character formulas corresponding to \eqn{group 1} are summarized in \cite{ES-G2orb}
(based on \cite{Dobrev,ZF2,Qiu,RY2}): 
\begin{eqnarray}
\chi^k_{\ell \,[0,1]}(\tau)& =& 
\left\{
\begin{array}{ll}
\dsp  \frac{2}{\th_2(\tau)} \left(
\Th{2(\ell+1)}{4(k+2)}(\tau)+(-1)^k\Th{2(\ell+1)+4(k+2)}{4(k+2)}(\tau)
\right)   &  ~~ (\ell~:~\mbox{even}) ,\\
0 & ~~(\ell~:~\mbox{odd}).
\end{array}
\right.
\nn
\chi^k_{\ell\,[1,0]}(\tau)&=& \frac{1}{\th_4(\tau)}
\,\left(\Th{\ell+1-\frac{k+2}{2}}{k+2}(\tau)-
\Th{-(\ell+1)-\frac{k+2}{2}}{k+2}(\tau)\right)  \nn
&= & \frac{1}{\th_4(\tau)}\left(\Th{2(\ell+1)-(k+2)}{4(k+2)}(\tau)
 + \Th{2(\ell+1)+3(k+2)}{4(k+2)}(\tau) \right. \nn
 && \left. - \Th{-2(\ell+1)-(k+2)}{4(k+2)}(\tau)
 - \Th{-2(\ell+1)+3(k+2)}{4(k+2)}(\tau) \right)~,
\nn
\chi^k_{\ell\,[1,1]}(\tau)&=& 
\frac{1}{\th_3(\tau)}
\left(\Th{2(\ell+1)-(k+2)}{4(k+2)}(\tau)
 +(-1)^k \Th{2(\ell+1)+3(k+2)}{4(k+2)}(\tau) \right. \nonumber \\
 && \left. +(-1)^{\ell} \Th{-2(\ell+1)-(k+2)}{4(k+2)}(\tau)
 +(-1)^{k+\ell} \Th{-2(\ell+1)+3(k+2)}{4(k+2)}(\tau) \right)~.
\label{twisted minimal}
\end{eqnarray}
The conformal dimensions of the ground states corresponding to the first characters are
\begin{eqnarray}
h= h_{\ell} \equiv \frac{\ell(\ell+2)}{4(k+2)}~,
\end{eqnarray}
(which coincide with those for the $SU(2)_k$ primaries).
The ground states of the second and third ones have dimensions 
\begin{eqnarray}
h= h_{\ell}^{t} \equiv  
\frac{k-2+(k-2\ell)^2}{16(k+2)}+\frac{1}{16}~.
\label{h t}
\end{eqnarray}
The states characterised by \eqn{h t} are interpreted as the product of 
the twist field in the $U(1)$-sector and the ``$C$-disorder field'' \cite{ZF2} 
in the $\bz_k$-parafermion theory \cite{ZF1}. 
Note that
$\chi^k_{k-\ell\,[1,0]}=\chi^k_{\ell\,[1,0]}$, 
$\chi^k_{k-\ell\,[1,1]}=\chi^k_{\ell\,[1,1]}$.
Due to these relations the corresponding fields are identified,
leaving only $\ell=0, 1,\ldots, \left\lb \frac{k}{2}\right\rb$ as 
independent primary fields.

The modular transformations of the twisted $\cN =2$ characters are 
\begin{eqnarray}
&&\hskip-15mm
\chi^k_{\ell\,[0,1]}(\tau+1)= 
e^{2\pi i \left(h_\ell -\frac{k}{8(k+2)}\right)}\,
\chi^k_{\ell\,[0,1]}(\tau)~, 
\hskip5mm
\chi^k_{\ell\,[0,1]}\left(-\frac{1}{\tau}\right)
=\sum_{\ell'=0}^k\, (-1)^{\ell/2} 
S_{\ell,\ell'}\,
\chi^k_{\ell'\,[1,0]}(\tau),  \nn
&&\hskip-15mm
\chi^k_{\ell\,[1,0]}(\tau+1)=
e^{2\pi i\left(h^t_\ell-\frac{k}{8(k+2)}\right)}\, 
\chi^k_{\ell\,[1,1]}(\tau)~, 
\hskip5mm
\chi^k_{\ell\,[1,0]}\left(-\frac{1}{\tau}\right)=
\sum_{\ell'=0}^k\, S_{\ell,\ell'}(-1)^{\ell'/2}\, 
\chi^k_{\ell'\,[0,1]}(\tau)~, \nn
&&\hskip-15mm \chi_{\ell\,[1,1]}(\tau+1)= 
e^{2\pi i \left(h^t_\ell -\frac{k}{8(k+2)}\right)}\,
\chi^k_{\ell\,[1,0]}(\tau)  ~, 
\hskip5mm
\chi^k_{\ell\,[1,1]}\left(-\frac{1}{\tau}\right)=
\sum_{\ell'=0}^k\, \widehat{S}_{\ell,\ell'}\,
\chi^k_{\ell'\,[1,1]}(\tau) ~ .
\label{modular twisted minimal} 
\end{eqnarray}
Here 
$ S_{\ell,\ell'}\equiv \sqrt{\frac{2}{k+2}}
\sin\left(\frac{\pi(\ell+1)(\ell'+1)}{k+2}\right)$ is 
the modular S-matrix of the $SU(2)$ WZW model at level $k$, 
and 
$\widehat{S}_{\ell,\ell'}\equiv e^{\frac{\pi i}{2}\left(\ell+\ell'-\frac{k}{2}\right)}\, 
S_{\ell,\ell'}$.

Finally, we mention the remaining minimal model characters appearing in 
\eqn{group 2}. 
In contrast to the $\hat{c}>1$ case, these characters always vanish.
For instance, let us pick up the boundary condition $\{\R,\, [0,1]\}$. 
Only the representations generated by doubly degenerated primary states
$\ket{\ell,m,s}=\ket{\ell,0,\pm 1}$ ($\ell \in 2\bz+1$) 
can contribute, but the trace over them vanishes because 
$\sigma^{\cN=2}$ acts as 
$$
\sigma^{\cN=2}~:~ \ket{\ell,0,\pm 1}~\longmapsto ~ \ket{\ell,0,\mp 1}~.
$$

~


\section*{Appendix C ~ Twisted $SU(2)_k$ Characters}
\setcounter{equation}{0}
\def\theequation{C.\arabic{equation}}

The twisted characters of the $SU(2)_k$ current algebra are generally  
written as
\begin{eqnarray}
 && \chi^{SU(2)_k}_{\ell,(a,b)}(\tau,z)
\equiv e^{2\pi i \frac{k}{4}ab} q^{\frac{k}{4}a^2}
y^{\frac{k}{2}a}\, \chi^{SU(2)_k}_{\ell}(\tau,z+a\tau+b)~,
\label{twisted SU(2) ch}
\end{eqnarray}
where $a$ and $b$ parameterize the spatial and temporal boundary
conditions. This is a special case of more general formulas for the
twisted characters of affine Kac-Moody algebras \cite{BFS} (up to
phase factors). 
Especially, $(a,b)=(0,1/2)$ corresponds to temporal insertion of 
$e^{i\pi J^3_0}$ within the trace and by direct calculations we may show that
it is related to the twisted $\cN =2$ characters by\footnote
   {Since $\chi^{SU(2)_k}_{\ell,(a/2,b/2)}(\tau,0)$ 
   and $\chi^k_{\ell,[[a],[b]]}(\tau)$ ($[a]\in \bz_2$ is defined 
   by $a\equiv [a]~ \mod\, 2$)
   differ only by a phase factor,
   $\chi^k_{\ell,[[a],[b]]}(\tau)$ may also be regarded as twisted 
   $SU(2)_k$ characters. 
   In fact, the same character functions $\chi^k_{\ell,[[a],[b]]}(\tau)$ are employed
   in \cite{KawaiS1} to analyse twisted representations that display
   manifest $\bz_2$-periodicities in the twist parameters. 
   In that paper, formulas involving the angular variable dependence ($z$) 
   associated with the $SU(2)$ zero-modes are presented. 
   In the $\cN=2$ case the $z$-dependence is irrelevant because 
   the $\cN=2$-involution $\sigma^{\cN=2}$ removes the zero-mode 
  of the $U(1)$-current $J$. 
},
\begin{eqnarray}
 && \chi^{SU(2)_k}_{\ell,(0,1/2)}(\tau,0) 
= (-1)^{\ell/2} \chi^k_{\ell,[0,1]}(\tau)~.
\label{rel twisted N=2 SU(2)}
\end{eqnarray}
(Both $\chi^{SU(2)_k}_{\ell,(0,1/2)}(\tau,0)$ and 
$\chi^k_{\ell,[0,1]}(\tau)$ vanish when $\ell$ is odd,
so the factor $(-1)^{\ell/2}$ entails no phase ambiguity.)
Performing modular transformations, 
we further obtain
\begin{eqnarray}
 && \chi^{SU(2)_k}_{\ell,(1/2,0)}(\tau,0) 
= \chi^k_{\ell,[1,0]}(\tau)~, \nn
&& \chi^{SU(2)_k}_{\ell,(1/2,1/2)}(\tau,0) 
= e^{2\pi i \frac{k}{16}} e^{-\frac{i\pi}{2}\ell}\chi^k_{\ell,[1,1]}(\tau)~.
\label{rel twisted N=2 SU(2) 2}
\end{eqnarray}

The modular property of the twisted character $\chi^{SU(2)_k}_{\ell,(a,b)}(\tau,z)$ is simply
written as
\begin{eqnarray}
 && \chi^{SU(2)_k}_{\ell,(a,b)}\left(-1/\tau,z/\tau\right)
= e^{i\pi \frac{k}{2}\frac{z^2}{\tau}} \, \sum_{\ell'=0}^k \,
S^{}_{\ell,\ell'}\,
\chi^{SU(2)_k}_{\ell',(b,-a)}\left(\tau,z\right)~,
\label{twisted SU(2) S} \\
&& \chi^{SU(2)_k}_{\ell,(a,b)}(\tau+1,z) =
e^{2\pi i \left(\frac{\ell(\ell+2)}{4(k+2)}-\frac{k}{8(k+2)}\right)}\,
\chi^{SU(2)_k}_{\ell,(a,a+b)} (\tau,z)~.
\label{twisted SU(2) T}
\end{eqnarray}

~


\section*{Appendix D ~ Complete Classification of Twisted $\cN=4$ Characters}
\setcounter{equation}{0}
\def\theequation{D.\arabic{equation}}

In this appendix we present a complete classification of the twisted
$\cN=4$ characters. 

~

\noindent
{\bf 1. $\sigma^{\cN=4}_1$-twist : }

As we already discussed, a major part of the $\sigma^{\cN=4}_1$-twisted 
characters are exhibited in \eqn{twisted N=4 characters}. We thus focus
on the remaining sectors. It is enough to consider the 
$\{\R,\,[0,1]\}$ ($\{\tR,\,[0,1]\}$) sector, which is the trace over each
Ramond representation with $\sigma^{\cN=4}_1$ ($(-1)^F
\sigma^{\cN=4}_1$) inserted.
The remaining ones 
$\{\tNS,\,[1,0]\}$, $\{\NS,\,[1,1]\}$ ($\{\tR,\,[1,0]\}$, $\{\tR,\,[1,1]\}$)
are generated by modular transformations. As already mentioned, 
$\sigma^{\cN=4}_1$ boils down to the $\sigma^{\cN=2}$-twist and 
the spectral-flowed sectors do not contribute. 
Thus we find 
\begin{eqnarray}
&& \tr_{\cC^{(\sR)}_h} \left[ \sigma^{\cN=4}_1 q^{L_0-\frac{1}{4}}\right]
 =\tr_{\cD^{(\sR)}_{1/2}} \left[ \sigma^{\cN=4}_1
			   q^{L_0-\frac{1}{4}}\right]
=0~, ~~~ \tr_{\cD^{(\sR)}_{0}} \left[ \sigma^{\cN=4}_1 
q^{L_0-\frac{1}{4}}\right] = \pm 1~. 
\label{sigma 1 R 0 1 ch}
\end{eqnarray}
We also obtain the same results for the $\{\tR,\,[0,1]\}$-characters. 
It is trivial to modular transform these results to obtain the remaining ones.

~


\noindent
{\bf 2. $\sigma^{\cN=4}_3$-twist : }

The equivalence of twisted character formulas for $\sigma^{\cN=4}_3$ and 
$\sigma^{\cN=4}_1$ is anticipated; we shall verify this explicitly. 

Again we focus on the $\{\R,\,[0,1]\}$ and $\{\tR,\,[0,1]\}$ sectors,
since the classification \eqn{twisted N=4 characters} has been already
given. Namely, we examine the trace over each Ramond representation with 
the insertion of $\sigma^{\cN=4}_3$ (and 
$(-1)^F \sigma^{\cN=4}_3$), which assigns the phase $(-1)^n$ to 
the $n$-th spectral flow sector. 
For representations $\cC^{(\sR)}_{h}$, $\cD^{(\sR)}_{1/2}$, 
($\cC^{(\stR)}_{h}$, $\cD^{(\stR)}_{1/2}$)
we readily obtain 
\begin{eqnarray}
 && \tr_{\cC_h^{(\sR)}}\left[\sigma^{\cN=4}_3 q^{L_0-\frac{1}{4}}\right]
(\equiv  \tr_{\cC_h^{(\sR)}}\left[(-1)^{F}\sigma^{\cN=4}_3 q^{L_0-\frac{1}{4}}\right])
= q^{h-\frac{3}{8}} \frac{i\th_1(\tau,0)\th_2(\tau,0)}{\eta(\tau)^3}
=0~, \\
 && \tr_{\cD_{1/2}^{(\sR)}}\left[\sigma^{\cN=4}_3 q^{L_0-\frac{1}{4}}\right]
(\equiv  \tr_{\cD_{1/2}^{(\sR)}}
\left[(-1)^{F}\sigma^{\cN=4}_3 q^{L_0-\frac{1}{4}}\right])
= q^{-\frac{1}{8}} \frac{i\th_1(\tau,0)\th_2(\tau,0)}{\eta(\tau)^3}
=0~,
\label{sigma 3 R 0 1 ch}
\end{eqnarray}
by using \eqn{twisted N=4 massive}, \eqn{twisted N=4 grav} and the
1/2-spectral flow.
The one for the representation $\cD^{(\sR)}_{0}$ is somewhat
non-trivial:
\begin{eqnarray}
 && \tr_{\cD_0^{(\sR)}}\left[\sigma^{\cN=4}_3 q^{L_0-\frac{1}{4}}\right]
= \pm \sum_{n\in \bsz} (-1)^n
\frac{q^{\frac{1}{2}n(n+1)}}
{1+q^n} \frac{\th_2(\tau)}{\eta(\tau)^3} \equiv \pm 1~.
\label{sigma 3 R 0 1 ch-2}
\end{eqnarray}
(Again we include a sign ambiguity due to the $\sigma_3^{\cN=4}$-action
on the vacuum.) 
The second equality follows from the identity (e.g. (3.17) in \cite{KacW})
\begin{eqnarray}
&&  \frac{1}{\prod_{n=1}^{\infty}(1+yq^{n-\frac{1}{2}})(1+y^{-1}q^{n-\frac{1}{2}})}
= \frac{q^{\frac{1}{12}}}{\eta(\tau)^2}\, \sum_{n\in \bsz} (-1)^n
\frac{q^{\frac{1}{2}n(n+1)}}{1+yq^{n+\frac{1}{2}}}~,
\label{KW identity}
\end{eqnarray}
which can be derived from the super boson-fermion correspondence
\cite{KacW,Kac}.\footnote
  {More general identities for the level $k$ Appell function 
$
\cK_k(\tau,\nu,\mu)\equiv \sum_{n\in\bsz}\frac{q^{\frac{k}{2}n^2}x^{kn}}{1-xyq^n}
$, ($x\equiv e^{2\pi i \nu}$, $y\equiv e^{2\pi i \mu}$)
are given in \cite{Pol,STT}. 
The $k=1$ case \cite{Pol}
is relevant here:
$$
\th_3(\tau,\la) \cK_1(\tau,\nu,\mu) -
\th_3(\tau,\nu) \cK_1(\tau,\la,\mu)
= i \frac{\th_3(\tau,\nu+\mu+\la)\th_1(\tau,-\nu+\la)}
{\th_1(\tau,\nu+\mu)\th_1(\tau,\mu+\la)} \eta(\tau)^3~,
$$
from which one can reproduce 
the identity \eqn{KW identity} by setting $\la=\frac{\tau+1}{2}$.
Generalization to higher level cases has been given in \cite{STT}
(Lemma 2.2). 
}

On the other hand, if making the $(-1)^F \sigma^{\cN=4}_3$-insertion, 
only the Ramond ground states can contribute, and it is easy to see 
\begin{eqnarray}
 && \tr_{\cD_0^{(\sR)}}\left[(-1)^F\sigma^{\cN=4}_3 q^{L_0-\frac{1}{4}}\right]
= \pm 1~.
\label{sigma 3 R 0 1 ch-3}
\end{eqnarray}

In this way we have confirmed the equality of the $\sigma_1^{\cN=4}$-
and $\sigma_3^{\cN=4}$-twisted characters for all the 
irreducible representations of $\cN=4$ SCA.



\end{document}